\RequirePackage{scrlfile}
\PreventPackageFromLoading{subfig}
\documentclass[twoside,leqno,twocolumn]{article}

\usepackage[letterpaper]{geometry}
\usepackage{ltexpprt}

\usepackage{cite}
\usepackage{programs}
\usepackage{hyperref}
\usepackage{graphicx}
\usepackage{svg}
\usepackage{hyperref}
\usepackage{subcaption}
\graphicspath{{images/}}

\begin{document}

\title{\Large Lock-Free Hopscotch Hashing}
\author{
Robert Kelly\thanks{Maynooth University Department of Computer Science. Email: \href{mailto:rob.kelly@cs.nuim.ie}{rob.kelly@cs.nuim.ie}}
\and Barak A. Pearlmutter\thanks{Maynooth University Department of Computer Science.
Email: \href{mailto:barak@cs.nuim.ie}{barak@cs.nuim.ie}}
\and Phil Maguire\thanks{Maynooth University Department of Computer Science. Email: \href{mailto:pmaguire@cs.nuim.ie}{pmaguire@cs.nuim.ie}}
}

\date{}

\maketitle


\fancyfoot[R]{\scriptsize{Copyright \textcopyright\ 2020 by SIAM\\
Unauthorized reproduction of this article is prohibited}}

\begin{abstract} \small\baselineskip=9pt
In this paper we present a lock-free version of Hopscotch Hashing. Hopscotch Hashing is an open addressing algorithm originally proposed by Herlihy, Shavit, and Tzafrir \cite{Hopper}, which is known for fast performance and excellent cache locality. The algorithm allows users of the table to skip or jump over irrelevant entries, allowing quick search, insertion, and removal of entries. Unlike traditional linear probing, Hopscotch Hashing is capable of operating under a high load factor, as probe counts remain small. Our lock-free version improves on both speed, cache locality, and progress guarantees of the original, being a chimera of two concurrent hash tables. We compare our data structure to various other lock-free and blocking hashing algorithms and show that its performance is in many cases superior to existing strategies. The proposed lock-free version overcomes some of the drawbacks associated with the original blocking version, leading to a substantial boost in scalability while maintaining attractive features like physical deletion or \textit{probe-chain} compression.
\end{abstract}


\section{Introduction}
The trend in modern hardware development has shifted away from enhancing serial hardware performance towards multi-core processing. This trend forces programmers and algorithm designers to shift their thinking to a parallel mindset when writing code and designing their algorithms. Concurrent algorithms tackle the problem of sharing data and keeping that data coherent while multiple actors simultaneously attempt to change or access the data. For concurrent data structures and algorithms to perform well on modern processors they must generally have two properties. First, they must use the processor's cache memory efficiently for both data and instruction. Today's processors are very sensitive to memory access patterns and contention induced by concurrency in cache coherence protocols. As such, algorithm designers must take special care to accommodate these particulars. Second, the algorithms must ensure that reading the data structure is as cheap as possible. The majority of operations on structures like hash tables are read operations, meaning that it pays to optimise them for fast reads.

\begin{figure}[!htbp]
\centering
\includesvg[svgpath=./images/,width=.4\textwidth]{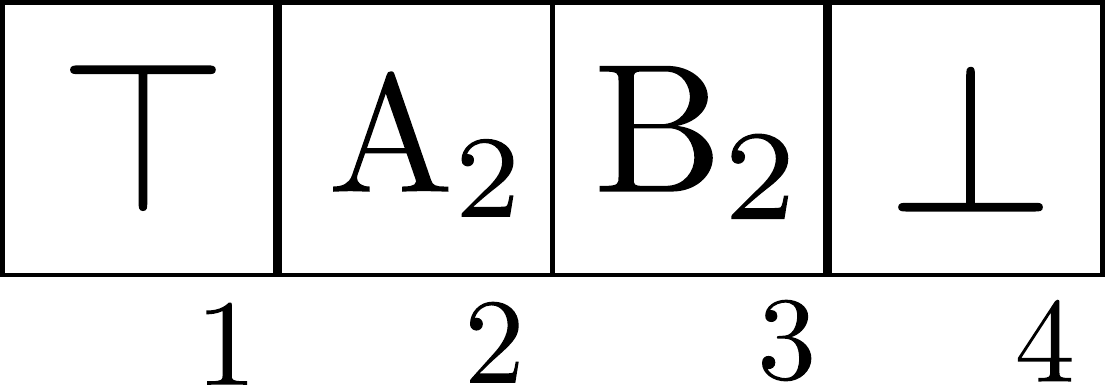}
\caption{Table legend. The subscripts represent the optimal bucket index for an entry, the $\top$ represents a tombstone, and the $\bot$ represents an empty bucket. Bucket 1 contains the symbol for a tombstone, bucket 2 contains an entry A which belongs at bucket 2, bucket 3 contains another entry B which also belongs at bucket 2, and bucket 4 is empty.}
\label{fig:Legend}
\end{figure}

Concurrent data structures and algorithms can be categorised into a variety of classes, with two prominent divisions emerging, namely \textit{blocking} and \textit{non-blocking}. Blocking algorithms generally involve the use of mutual exclusion primitives to give a single thread sole access to a specific location in memory. In contrast, non-blocking algorithms use low-level atomic primitives, such as \textit{compare-and-swap}, to modify the data structure. The category of non-blocking algorithms contains a number of further subdivisions. Ordered by the strength of progress these are: obstruction-free \cite{ObstructionFree} (individual progress in the case of no other contending operation), lock-free \cite{ampp} (system progress but not individual thread or actor progress), wait-free \cite{WaitFree} (every operation takes a finite number of steps to complete and must complete within that bound, which can depend on the number of threads or actors currently attempting such an operation), and wait-free population oblivious (every thread takes a finite amount of time regardless of how many other threads are in the system).

Maintaining strong progress conditions, or bounds on max latency, usually comes at the cost of throughput, meaning that the algorithms with the weakest guarantees typically boast the strongest real-world performance. Lock-free algorithms have the following benefits: freedom from deadlock, priority inversion, and convoying. However, they suffer from their own set of challenges relating to memory management (\cite{Alistarh:2015:TAS:2755573.2755600}, \cite{Alistarh:2017:FCM:3064176.3064214}, \cite{Michael:2004:HPS:987524.987595}, \cite{fraser2004practical}, \cite{cohen2015efficient}), proof of correctness (\cite{Herlihy:1990:LCC:78969.78972}, \cite{Leino2009}, \cite{DBLP:conf/pdp/AmighiBH16}) and poor performance under heavy write load, as excessive contention becomes an issue.

The outline of our paper is as follows. Section 2 gives the background for hash tables, a brief review of existing concurrent solutions, the original Hopscotch algorithm \cite{Hopper}, the Purcell-Harris quadratic probing hash table \cite{Purcell2005}, and the \textit{K-CAS} \cite{Harris:2002:PMC:645959.676137} primitive. Section 3 outlines our algorithm with code and annotations. Section 4 details our proof sketch. Finally section 5 discusses the experimental performance results, contrasting them with those of competing algorithms.

\section{Background}
A data structure for which concurrency is particularly amenable is the hash table. A hash table is an efficient base structure for implementing the abstract data structure of sets or maps. In general, hash tables compute a \textit{hash value} from a key or key-value pairing that a user seeks to either check membership, look up value, insert, or remove from the structure. The algorithm uses the hash value to index the location in which the entry should belong, and the entries are searched by following some strategy until a result is obtained. The expected lookup, insertion, and removal time bounds are $\mathcal{O}(1)$ \cite{Cormen:2001:IA:580470}. Entries need only be capable of being hashed to a location and compared for equality. In contrast, tree structures require a total ordering on keys or key/value pairings, but don't require a hash function.

Hash-tables are bifurcated into either open addressing or closed addressing algorithms. Open addressing constrains a bucket to contain only a single key or key-value pair. This constraint means that if two different keys or key-value pairings hash to the same index for an insertion, then an algorithm must be devised for finding another suitable bucket for insertion. The algorithm must then also be able to find the key or key-value pairing at some bucket outside of the original/home index. The alternative approach is closed addressing. Closed addressing stores all keys or key-value pairs at the original hash index. If two keys or key-value pairs collide at an address, then they are stored in an auxiliary data-structure like a linked-list or binary tree for searching. Closed addressing is therefore relatively simple and concise, needing only to search a single bucket when examining the table for an entry. Open addressing can be more challenging, as buckets contain entries which don't belong there but rather are there due to a previous collision. There has been many publications covering both concurrent open addressing \cite{GaoTable}, \cite{CliffClick}, \cite{Purcell2005}, \cite{Hopper}, \cite{NguyenCuckoo}, \cite{Nielsen:2016:SLH:3016078.2851196}, \cite{kelly_et_al:LIPIcs:2018:10070} and closed addressing algorithms \cite{Michael2002}, \cite{Shalev03}, \cite{WaitFreeExtensi}.

\begin{figure}[!htbp]
\centering
\includesvg[svgpath=./images/,width=.45\textwidth]{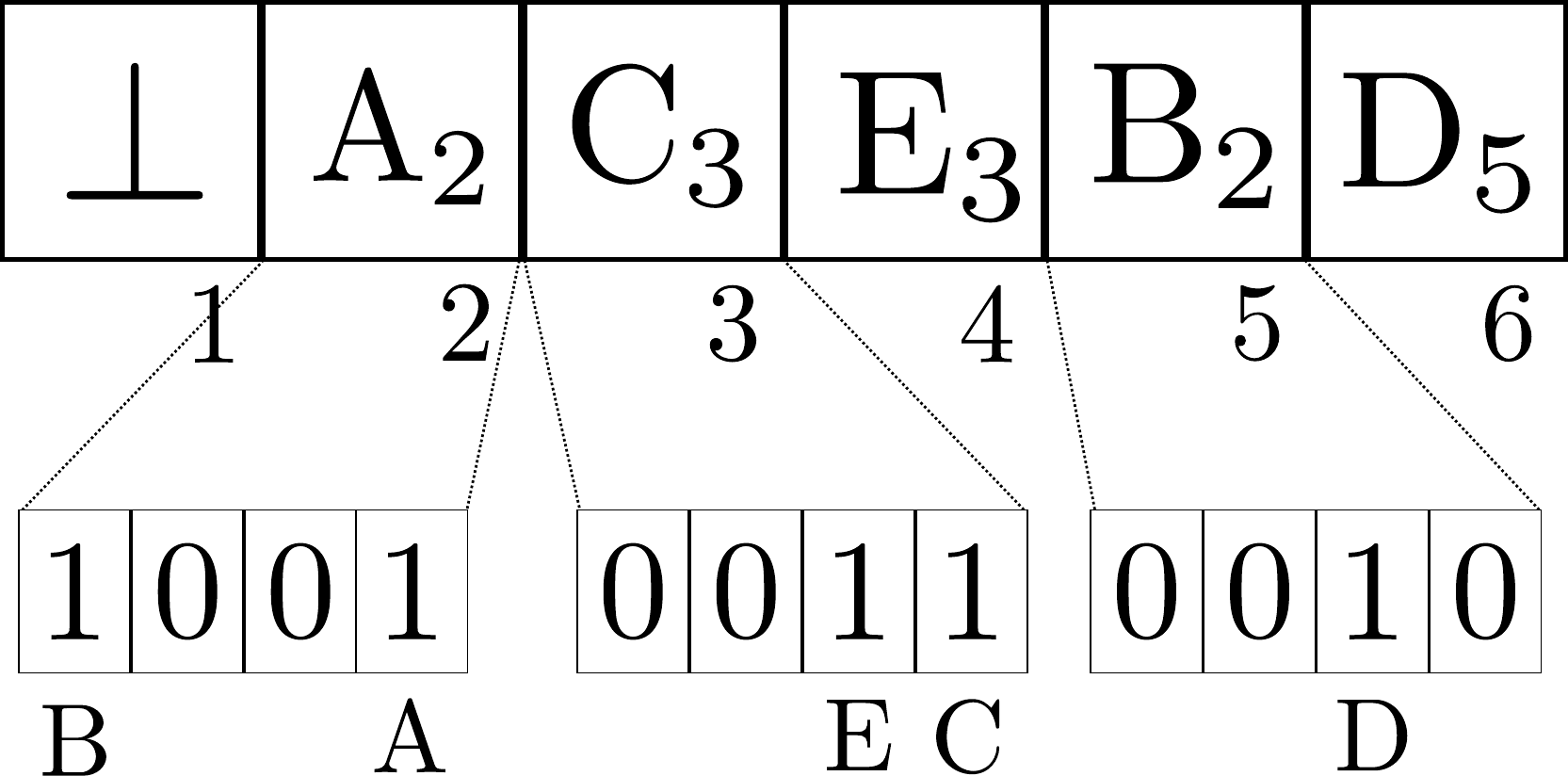}
\caption{An example Hopscotch Hashing table. The neighbourhood of virtual buckets is represented by a bit-map below each bucket. Each set bit represents the offset of buckets to examine for key matches. Note the bit endianness of the bit-mask.}
\label{fig:HopscotchHashing}
\end{figure}


\begin{figure*}[!htbp]
    \centering
    \subcaptionbox{A Separate Chaining table. Entries hashed to a bucket are put into a linked list at the bucket. Buckets with nulls are denoted with a diagonal line.\label{fig:SeparateChaining}}
    {
    \begin{minipage}[c]{0.45\textwidth}
    \includesvg[svgpath=./images/,width=0.95\textwidth]{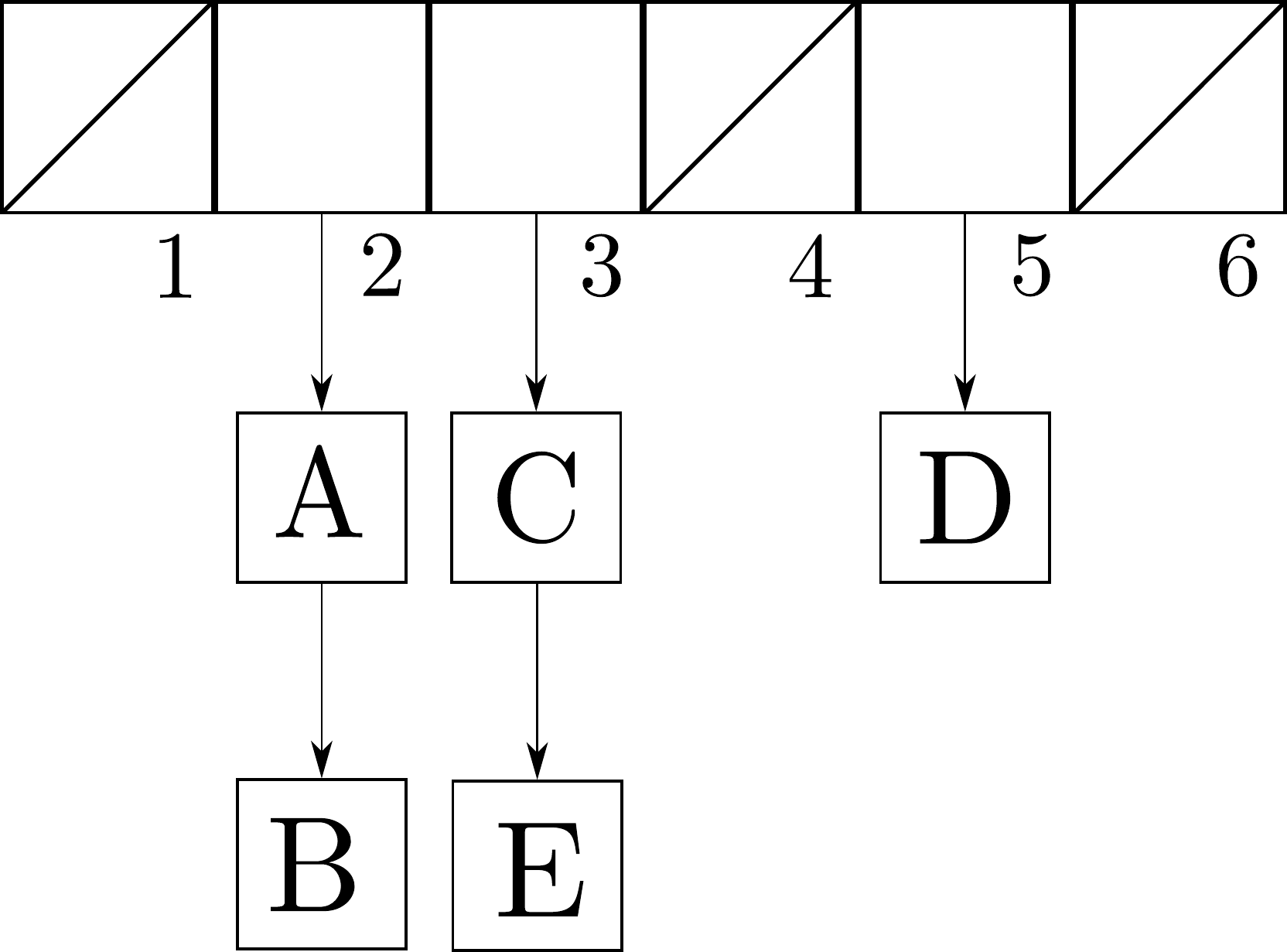}
    \end{minipage}
    }
    \qquad
    \subcaptionbox{A ``relative offset'' variant of Hopscotch Hashing table with the same entries as the Separate Chaining table. Each bucket contains two integers. The first is the offset where the probe chain starts, and the second is the next item in the probe chain. \label{fig:HopscotchRelativeOffset} }
    {
    \begin{minipage}[c]{0.45\textwidth}
    \includesvg[svgpath=./images/,width=0.95\textwidth]{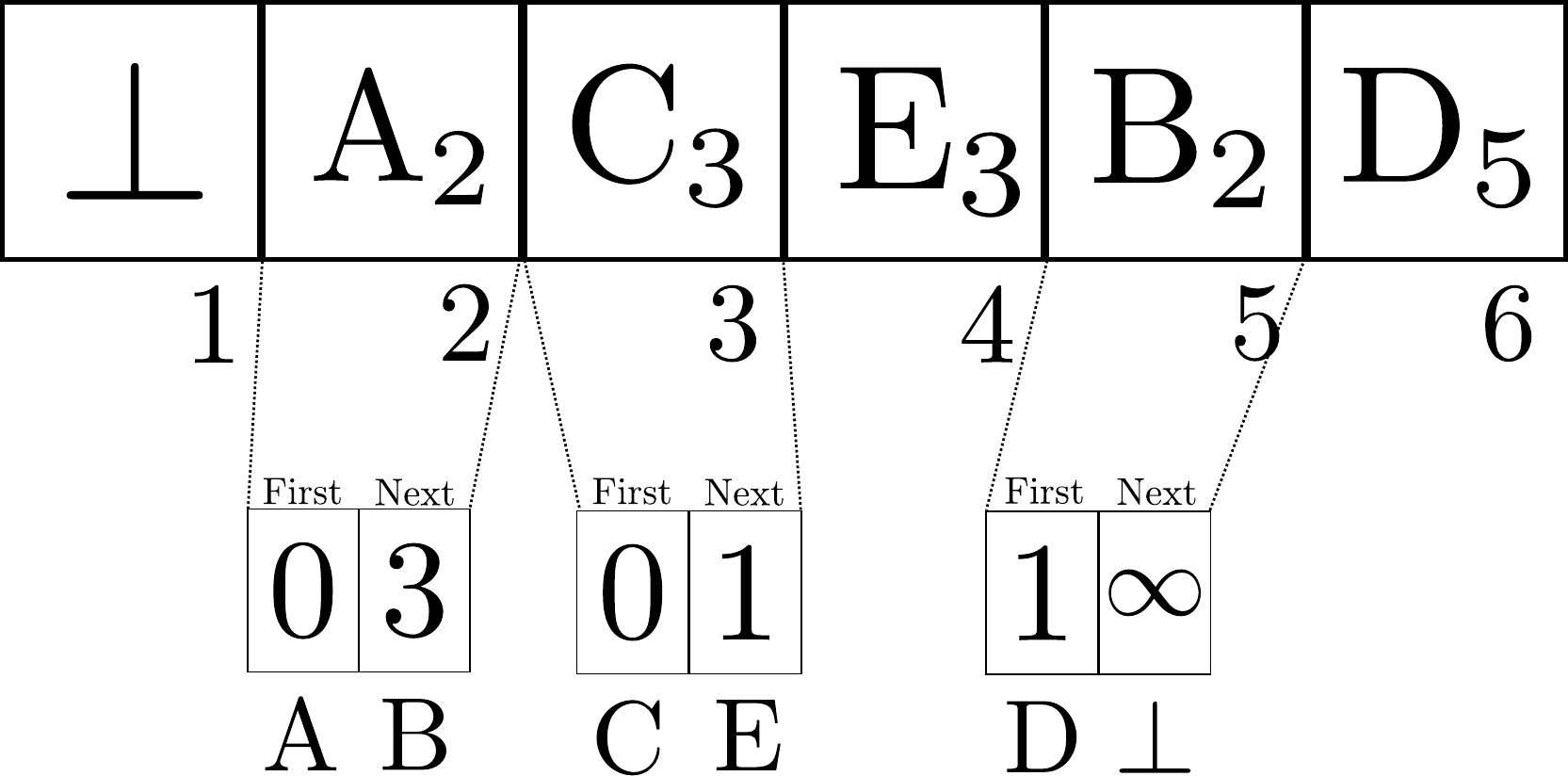}
    \vspace{1.8cm} 
    \end{minipage}
    }
    \caption{Comparison between Separate Chaining and Hopscotch Hashing with relative offsets.}
    \label{fig:Comparison}
\end{figure*}

\subsection{Original Hopscotch Hashing}
Herlihy, Shavit, and Tzafrir \cite{Hopper} presented Hopscotch Hashing, a hash table algorithm they describe as a mixture of linear probing \cite{LinearProbing}, separate chaining \cite{Knuth:1997:ACP:260999}, and Cuckoo Hashing \cite{SerialCuckoo}. Their paper presented solutions in both the serial and concurrent form. The algorithm comes in two main flavours. The first is to create a fixed sized \textit{neighbourhood} defining a virtual bucket containing many physical buckets. This neighbourhood is represented with a bit-mask, where the index of each set bit in the mask indicates the presence of a relevant entry for that particular bucket. An example table is shown in Figure \textbf{\ref{fig:HopscotchHashing}}, and a table legend is shown in Figure \textbf{\ref{fig:Legend}}. The algorithm solves collisions by linear probing to a free bucket and marking the $i$'th bit in the bit-mask, where $i$ is the distance from the original bucket. Due to the fixed size of the virtual buckets, a limit is enforced on how far an entry can be placed from its home/original bucket. The authors describe an algorithm for displacing entries from the saturated neighbourhood in order to create space for the new entry. The displacement algorithm works by linear probing to find an initial bucket and marking it as \texttt{Busy}. The algorithm then subsequently relocates the bucket backwards, swapping it with a currently occupied bucket, and modifying the occupied bucket's bit-mask during the move. Moving an occupied bucket forwards is only permissible if the destination bucket is also inside its neighbourhood. The algorithm repeats this process until the initially claimed bucket is within range of its home/original neighbourhood. If no such displacement can be made, then the table is forced to resize.

The authors then build upon this idea of the fixed size neighbourhood, using \textit{relative offsets} to indicate where the next relevant entry is stored. These offsets represent the hops throughout the table. Each entry is then part of a chain of entries, aptly named the \textit{probe chain}. The relative offsets, if large enough, can represent a neighbourhood as large as the table, removing the need to displace entries that otherwise would be outside the neighbourhood range. These hops, like the bit-masks, allow the method to skip over entries that are irrelevant to the search. For example, when a table using linear probing becomes saturated an entry may end up quite some distance from its original bucket. If such a situation were to arise in Hopscotch Hashing, then the last relevant entry to that original bucket would have the offset embedded into it, ``pointing'' to the new entry. The relative offset variant of Hopscotch can be thought of as a specialised version of Separate Chaining, in which the linked list present at each bucket has been flattened directly into the table. Figure \textbf{\ref{fig:Comparison}} illustrates a comparison between Separate Chaining and Hopscotch Hashing where each table has the same entries.

The use of relative offsets does not mean that the need to relocate entries disappears. The authors (in their released implementation) optimise the probe chain by shifting entries backwards when an entry earlier in the chain is removed. Resizes may still be required, as some entries may end up being further away from the last item in the probe chain than can be represented in the relative offsets. We choose the fixed size bit-mask as our model for our lock-free version of the algorithm. Their concurrent version employs mutual exclusion on threads wishing to insert or remove from the data structure, with remove operations incrementing a relocation counter relating to the relevant bucket deleted or moved. The reading thread will check the relocation counter before and after, to ensure that none of the entries have been shifted around during the reading. The number of segments is set to the expected concurrency exposure of the table.

\begin{figure}[!htbp]
\centering
\includesvg[svgpath=./images/,width=.45\textwidth]{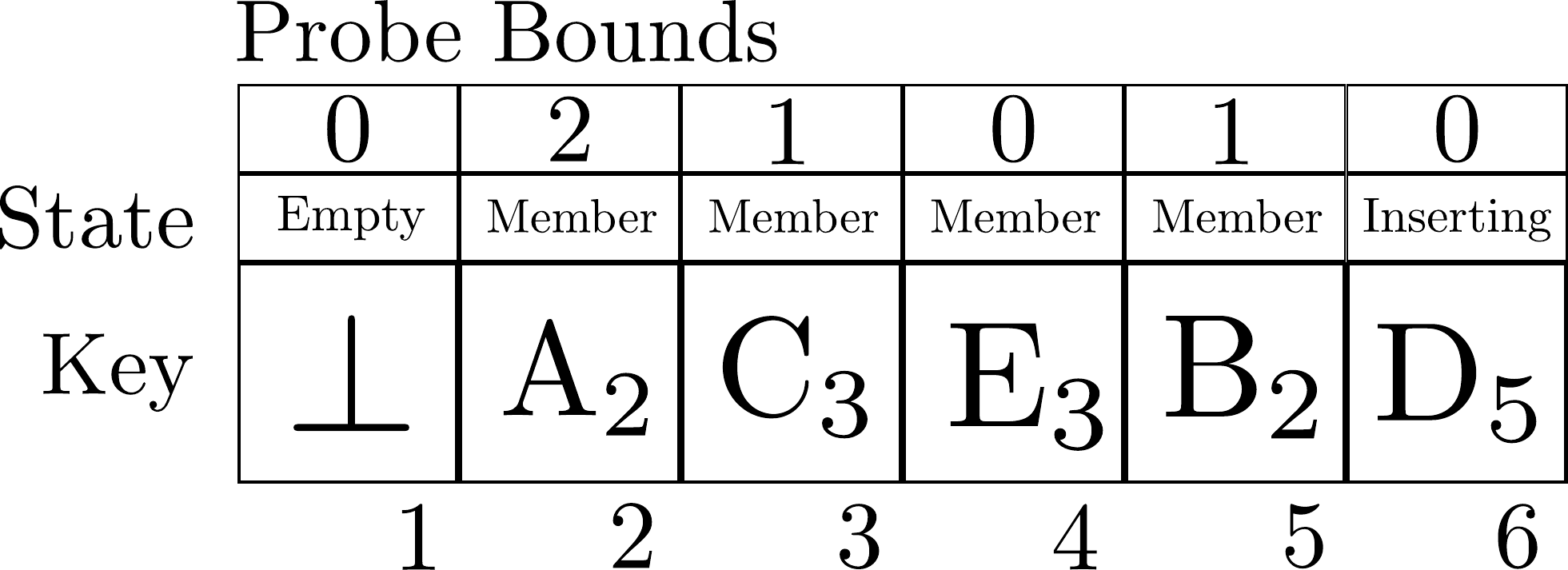}
\caption{An illustration of a Purcell-Harris table with bucket states and probe bounds.}
\label{fig:PurcellHarris}
\end{figure}

\subsection{Purcell-Harris Algorithm}
Our algorithm uses the Purcell-Harris method for insertion and deletion, so as to support physical deletion. Their approach uses a state based method for insertion and deletion. During insertion, keys or key-value pairs are eagerly inserted into the table and later checked for uniqueness. A bucket in their algorithm can be in 1 of 6 states, namely \texttt{Empty}, \texttt{Busy}, \texttt{Collided}, \texttt{Visible}, \texttt{Inserting}, or \texttt{Member}. \texttt{Empty} indicates that this bucket is empty and available for use in a new insertion. \texttt{Busy} can be seen as a \textit{lock}, used when the inserting/deleting thread is busy writing/deleting the key information in that bucket, and no one else may use the bucket. \texttt{Collided} is a state to indicate that the eager insertion of the entry has failed, as either a closer \textit{potential} entry exists, or an entry already marked as \texttt{Member} exists. \texttt{Visible} represents buckets which contain valid key data, but the bounds for per bucket probes may have not been updated yet. This allows other threads to see the entry and not decrease the probe length. The \texttt{Inserting} state means that the per bucket bounds have been updated to include this bucket and is in the process of being checked for uniqueness in the table. \texttt{Member} is the final stage of insertion, representing a unique key or key-value pair which is part of the table. All state variables contain an associated relocation counter to avoid the \textit{ABA} problem \cite{ABA_IBM}. An illustration of the table can be viewed in Figure \textbf{\ref{fig:PurcellHarris}}.

Removal of keys or key-value pairs is trivial, since they are atomically manipulated through the state variable. When removing an entry, the algorithm will simply move the state from \texttt{Member} to \texttt{Busy}, erase the key, potentially move the bound downwards, and then move the state back to \texttt{Empty}. Supporting physical deletion in non-blocking algorithms is difficult and is normally accomplished by putting each entry behind a dynamically allocated node. The Purcell-Harris algorithm makes this process trivial, while delivering good performance; all state variables and entries can be stored directly in the table, removing a level of indirection, and increasing cache efficiency. Checking for key membership is also straightforward. This simply involves reading the probe bound, examining any buckets marked as \texttt{Member} in the table, and checking that the associated version number hasn't changed since reading the state variable.

\subsection{K-CAS}
\textit{K-CAS} or, \textit{multi-word-compare-and-swap}, is an extension of the \textit{compare-and-swap} or, \textit{CAS} primitive. The algorithm allows for multiple memory locations to be atomically updated in the same fashion as a single \textit{CAS} operation. The quintessential algorithm for \textit{K-CAS} was published by Harris, Kaiser, and Pratt \cite{Harris:2002:PMC:645959.676137}, which works by installing shared operation \textit{descriptors} at each word being updated so that threads can cooperatively help each other complete the operation. To distinguish descriptors from normal words, up to 2 bits are reserved by the algorithm. For pointers, no extra bits are needed, as the lower bits are normally free; normal values like integers require 2 bits to be set aside. As a result of reserving bits, special read/write functions are needed when interfacing with memory. The read function checks the reserved bits of a word to see if any ongoing \textit{K-CAS} operation is currently in flight and, if so, assists in completing it.

The performance of \textit{K-CAS} was previously limited due to the necessity of a memory reclaimer. Each descriptor must be \textit{fresh} (newly allocated) to avoid the ABA problem \cite{ABA_IBM}, and, as such, the overhead was high. An algorithm by Arbel-Raviv and Brown \cite{ArbelRaviv2017ReuseDR} employs descriptor reuse, thereby eliminating the need for a freshly allocated descriptor for each operation. This substantially increases the performance of each \textit{K-CAS} operation, making their new algorithm a practical consideration when designing performant lock-free algorithms.

\section{Algorithm}
In the following section we provide an overview of our lock-free version of Hopscotch Hashing. The blueprint of our data-structure is that of a concurrent set, conforming to the API and abstract semantics. Our algorithm starts with the Purcell and Harris implementation of lock-free quadratic probing, and uses Hopscotch's bit-masks to create a fix-bound probe range for the searches. The check for uniqueness is then performed within that fix-sized area. The combination of the two algorithms removes the need for conditionally raising or lowering probe bounds, and allows for Hopscotch searching, insertion, and deletion in a lock-free manner. Like Purcell and Harris' quadratic probing, it allows for physical deletion, a difficult task to perform for a lock-free algorithm. We employ relocation counters at each bucket to indicate when that bucket's neighbourhood has experienced a bucket relocation, a necessity seeing as our algorithm moves entries around the table. All operations read the relocation counter before and after to ensure that no concurrent move operations have taken place, thus ensuring operation consistency. Our algorithm also makes use of \textit{multi-word-compare-and-swap} or \textit{K-CAS} \cite{Harris:2002:PMC:645959.676137} for an atomic swap. Previous work by Kelly, Pearlmutter, and Maguire \cite{kelly_et_al:LIPIcs:2018:10070} shows that an efficient algorithm for \textit{K-CAS} by Arbel-Raviv and Brown \cite{ArbelRaviv2017ReuseDR} is feasible in the construction of concurrent hash-tables.

\begin{figure}[!htbp]
\centering
\includesvg[svgpath=./images/,width=.45\textwidth]{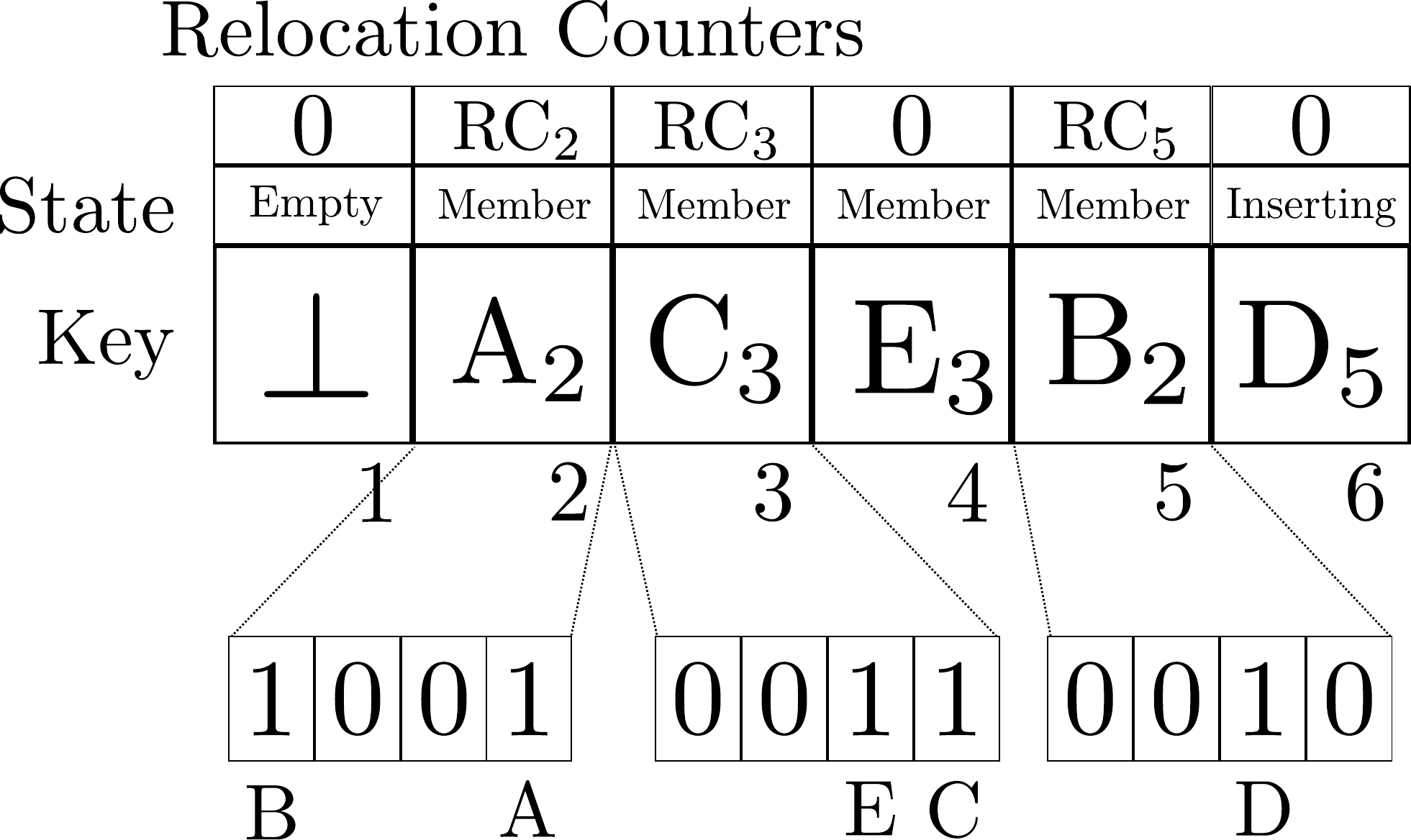}
\caption{An example table for lock-free Hopscotch Insertion. Our algorithm blends the Purcell-Harris state based buckets with Hopscotch bit-map neighbourhoods, fixing the probe bounds.}
\label{fig:HopscotchLockFree}
\end{figure}

When describing the table algorithm we start with the \textit{Add} method, as it influences the design of all other methods. The insertion process begins by checking if the key is already present in the table, reading the neighbourhood mask, and checking relevant buckets as indicated by each set bit. This check, however, is optional, as entries are inserted eagerly and checked for uniqueness afterwards. The algorithm then linear probes to find an \texttt{Empty} bucket and claims it, marking it as \texttt{Busy}. If the bucket claimed is within the neighbourhood range of the bit-mask, then a uniqueness check begins, completing the insertion. Otherwise the claimed bucket must be moved backwards towards its home neighbourhood bit-mask. To move a bucket back we use the standard Hopscotch displacement method, finding a suitable bucket to move forward, copying its key or key-value pair to the new bucket, marking it as \textit{live} in the neighbourhood bit-mask, and finally use \textit{K-CAS} to swap the bucket states and increment the moved bucket's relocation counter to force re-reads of the neighbourhood. If the \textit{K-CAS} is successful, we remove the bit previously corresponding to where the now moved entry previously resided. All of these searches also take account of ongoing relocations by reading relocation counters of buckets to ensure they don't miss a potential bucket. An illustration of the table is in Figure \textbf{\ref{fig:HopscotchLockFree}}.

\begin{figure*}[!htbp]
    \centering
    \subcaptionbox{Bucket \textbf{7} is claimed and marked as \texttt{Busy}. The bucket is outside the neighbourhood range and must be moved back. \label{fig:HopscotchInsert1}}
    {
    \begin{minipage}[c]{0.45\textwidth}
    \includesvg[svgpath=./images/,width=0.95\textwidth]{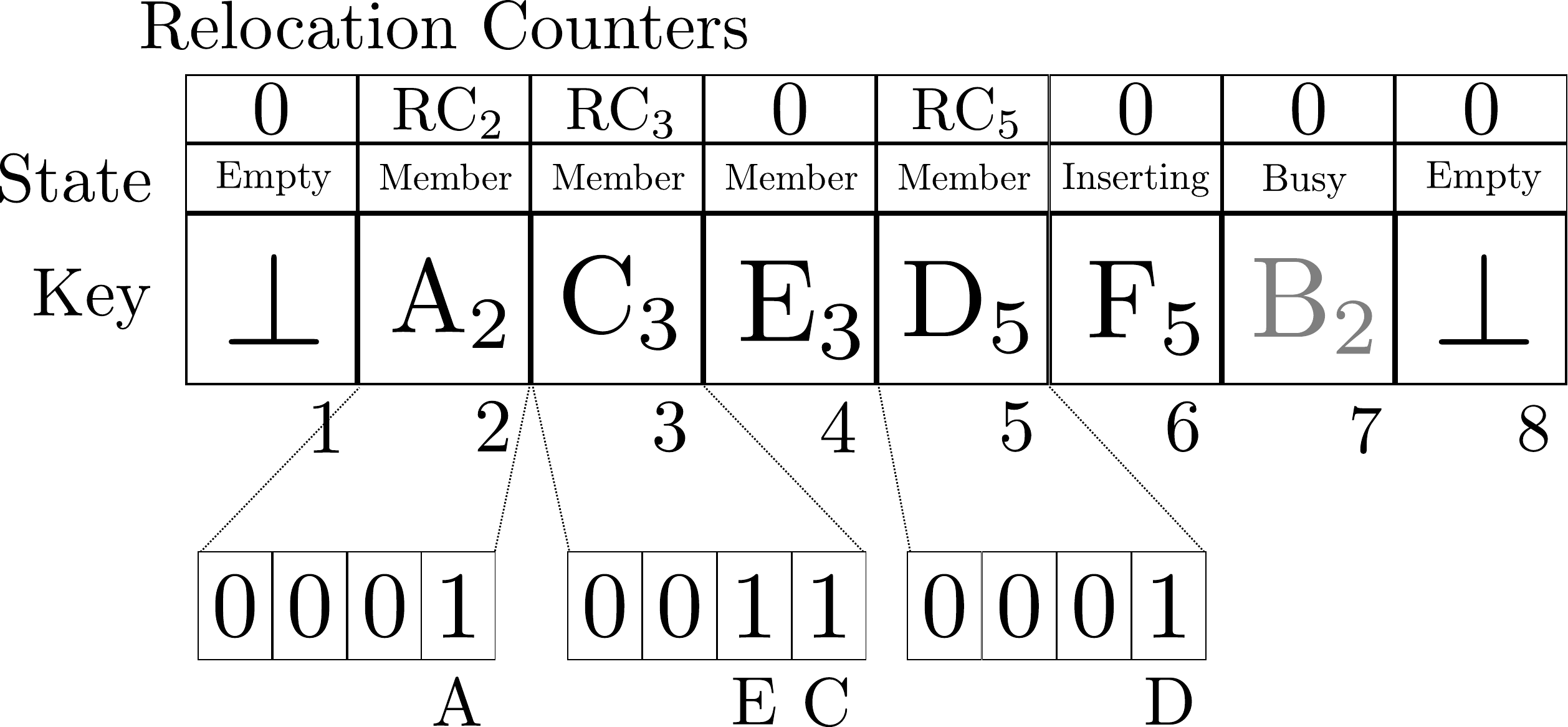}
    \end{minipage}
    }
    \qquad
    \subcaptionbox{Find a closer bucket and swap it with bucket 7, adding the bucket to the bit-mask and incrementing the relocation counter when swapping. \label{fig:HopscotchInsert2}
    }
    {
    \begin{minipage}[c]{0.45\textwidth}
    \includesvg[svgpath=./images/,width=0.95\textwidth]{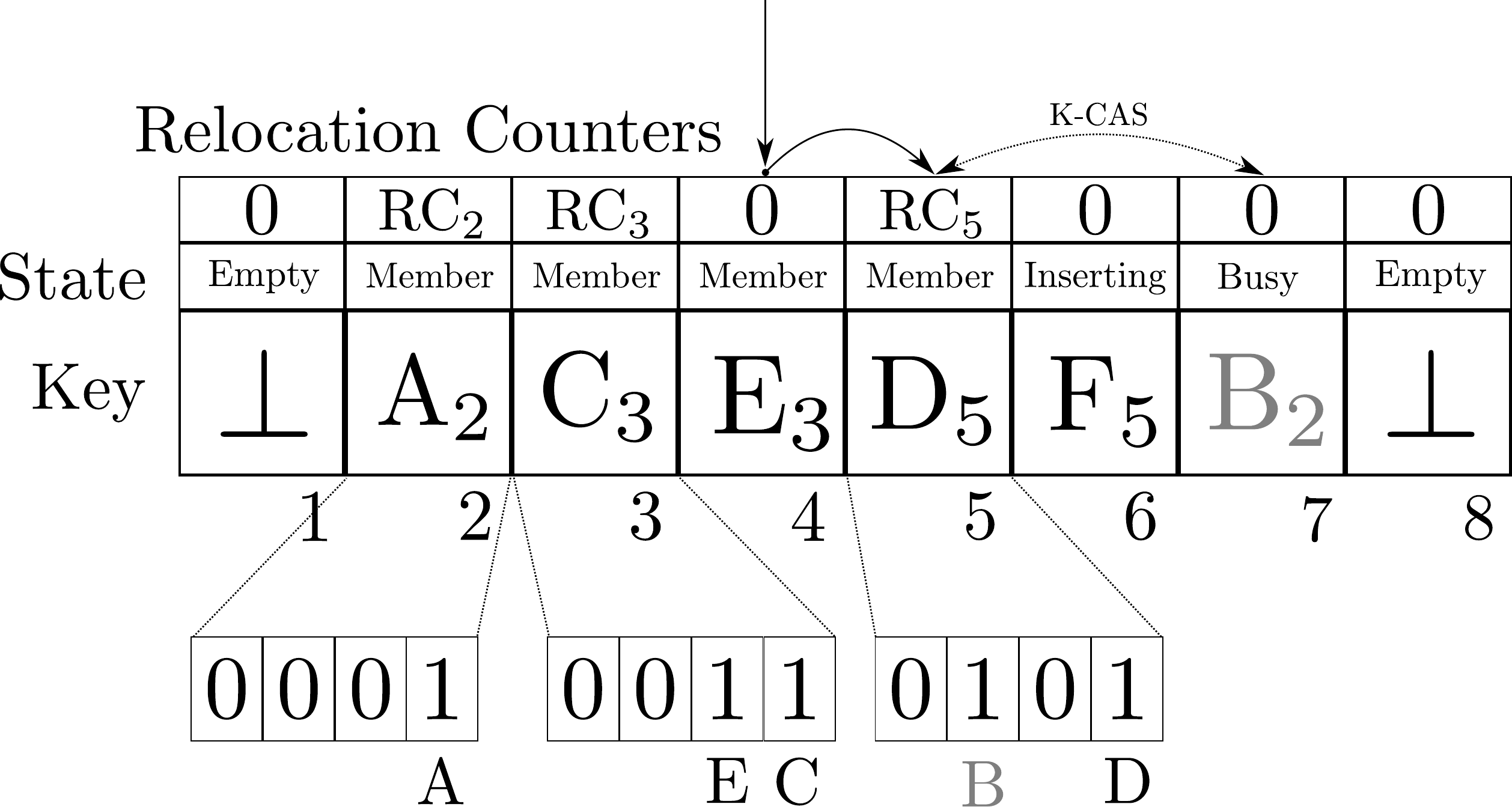}
    \end{minipage}
    }
    
    \subcaptionbox{Updated table after swapping buckets 5 and 7. The relocation counter at bucket 5 has been changed. \label{fig:HopscotchInsert3}}
    {
    \begin{minipage}[c]{0.45\textwidth}
    \includesvg[svgpath=./images/,width=0.95\textwidth]{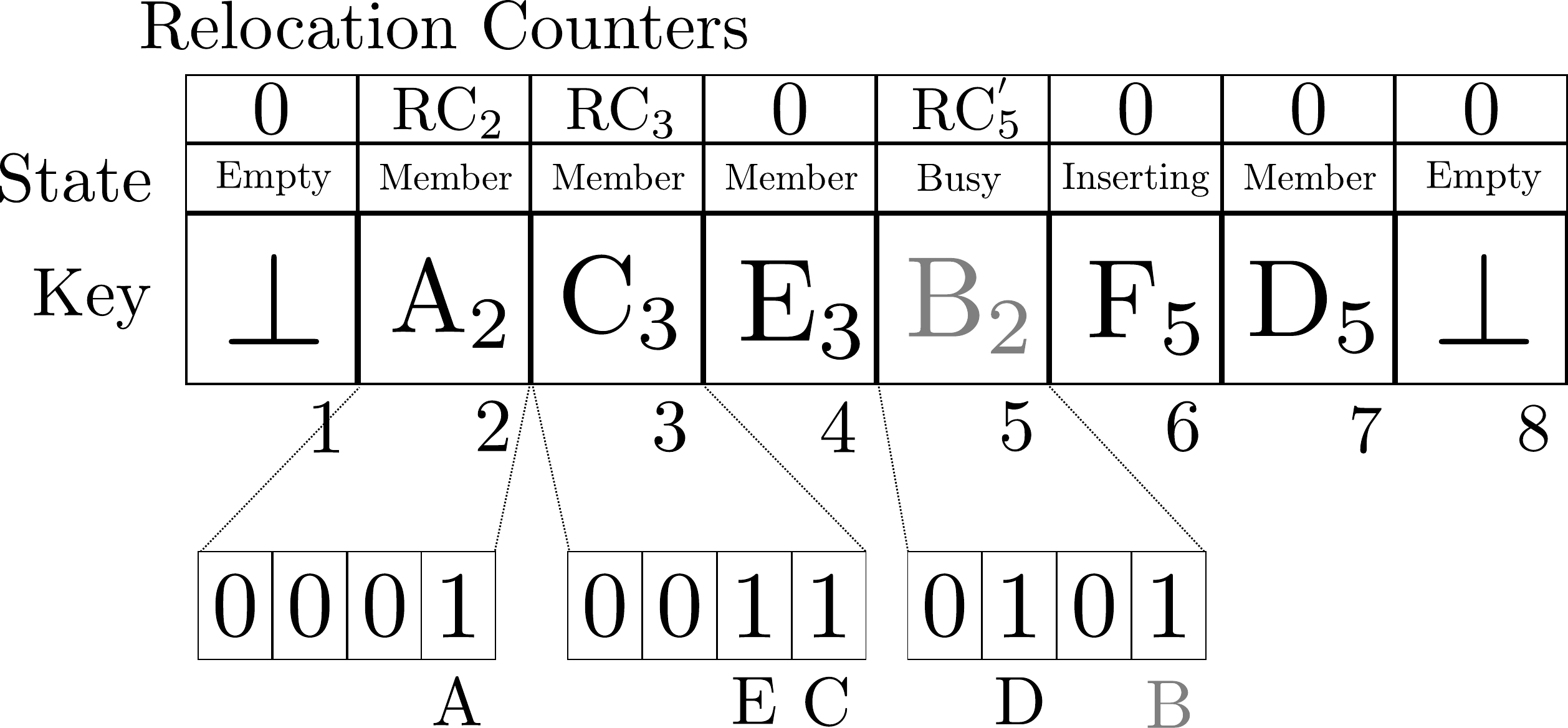}
    \vspace{0.27cm} 
    \end{minipage}
    }
    \qquad
    \subcaptionbox{Confirm bucket 5's uniqueness in the table, making the state of bucket 5 \texttt{Member} and adding it to bucket 2's bit-map. \label{fig:HopscotchInsert4}
    }
    {
    \begin{minipage}[c]{0.45\textwidth}
    \includesvg[svgpath=./images/,width=0.95\textwidth]{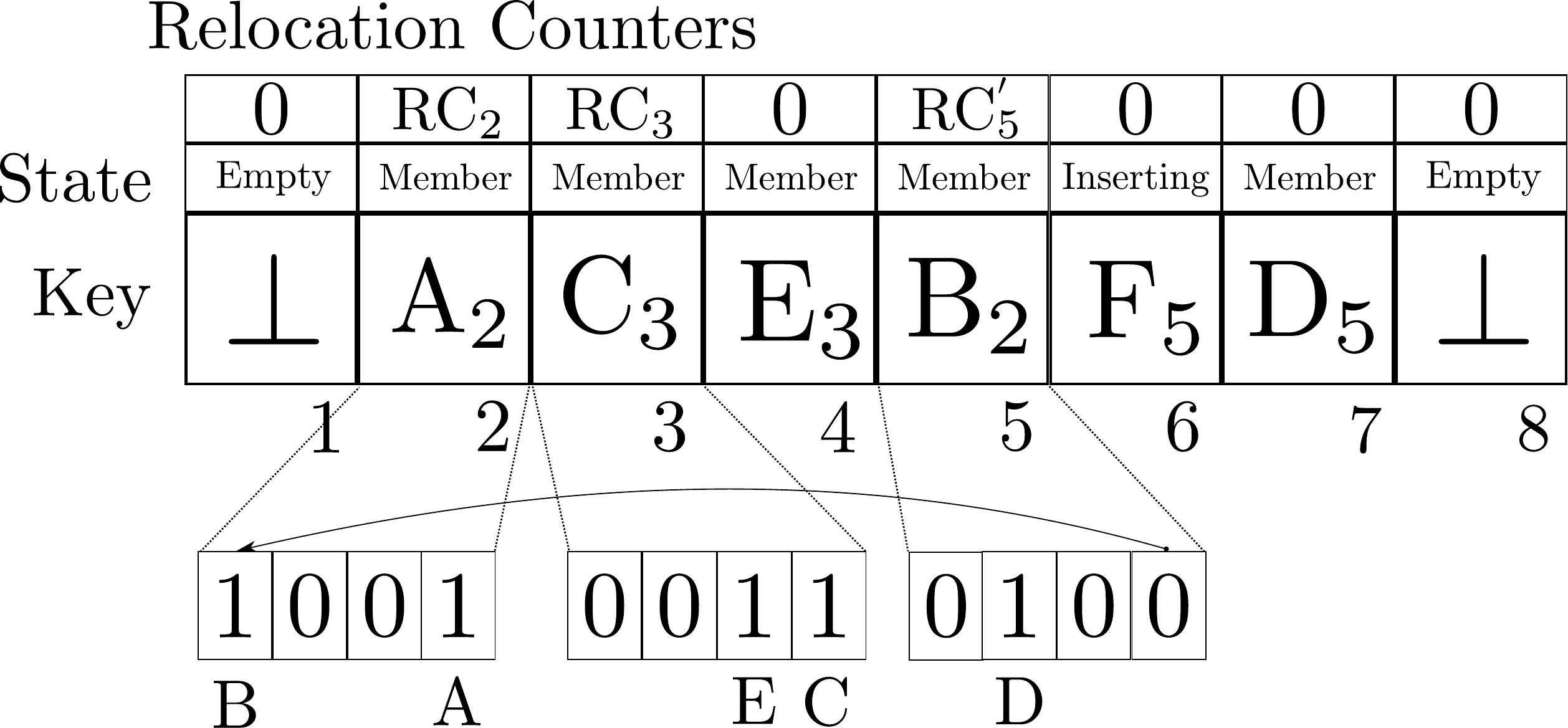}
    \end{minipage}
    }
    \caption{States of lock-free Hopscotch inserting element \texttt{B}.}
    \label{fig:HopscotchInsertion}
\end{figure*}

\begin{figure}
\begin{minipage}{0.01\textwidth}
\hspace{0.05cm}
\end{minipage}
\begin{minipage}{0.44\textwidth}
\begin{programf}
fn Contains(key: K) -> bool \{
  key\_hash = hash(key);
  ob = key\_hash \% size;
  rc\_before = table[ob].rc;
  while(true) \{
    // Load the neighbourhood bit-mask
    bm = table[ob].bm;
    while(bm != 0) \{
      // Find lowest set bit
      lsb = lowest\_set\_bit(bm);
      index = ob + lsb;
      // Purcell Harris bucket check.
      \_, state = table[index].vs;
      if(state == Member) \{
        if(table[index].k == key) \{
          return true;
        \}
      \}
      // Remove the bit just checked.
      bm = XOR(bm, 1 << lsb);
    \}
    // Check the relocation counter
    rc\_after = table[ob].rc;
    if(rc\_before == rc\_after) \{
      return false;
    \}
    // Check bit-mask again
    rc\_before = rc\_after;
  \}
\}
\end{programf}
\caption{Pseudo-code for \textit{Contains}}
\label{fig:Contains}
\end{minipage}
\end{figure}

Figure \textbf{\ref{fig:HopscotchInsertion}} gives a general overview of the insertion process, showing the many different stages for inserting an entry. Initially, claiming a bucket can be seen in Figure \textbf{\ref{fig:HopscotchInsert1}}. This bucket is outside the neighbourhood range and needs to be relocated backwards. The algorithm scans for potential buckets, as seen in Figure \textbf{\ref{fig:HopscotchInsert2}}, for a suitable entry to move. The first entry seen is entry \texttt{E}, which cannot be moved, as it would be outside its neighbourhood range. The next entry considered is \texttt{D}; moving it is legal. As the next entry, \texttt{B}'s location is marked in the bit-map for \texttt{D}, in anticipation of its movement. Following that, \textit{K-CAS} is employed to swap the two entries and update \texttt{D}'s relocation counter. The resulting configuration can be seen in Figure \textbf{\ref{fig:HopscotchInsert3}}, where \texttt{D} and \texttt{B} have been swapped and \texttt{D}'s relocation counter has been updated. The process continues as seen in Figure \textbf{\ref{fig:HopscotchInsert4}}. \texttt{B} removes itself from \texttt{D}'s bit-mask, and then checks if \texttt{B} is within its neighbourhood range, which it is. Lastly, the algorithm marks the bit in \texttt{B}'s home neighbourhood bit-mask, and, after adding it to the neighbourhood, the Purcell-Harris uniqueness check is performed within that neighbourhood, making entry \texttt{B} a member of the table.

\textit{Contains} and \textit{Remove} remain relatively simple, both loading the relocation counters of the neighbourhood being examined, along with the bit-mask. \textit{Contains} simply checks all buckets indicated by the bit-mask and successfully finishes if a key or key-value is found. If an entry isn't found, yet the relocation counter has changed, the method is performed again, returning an unsuccessful result if no change in the relocation counter is seen. \textit{Remove} follows the same process as \textit{Contains}, except once an matching entry is found it is put into a \texttt{Busy} state via a \textit{CAS}. If successfully put into a \texttt{Busy} state, then the key is removed, the relevant bit unset in the neighbourhood bit-mask, and the bucket marked as \texttt{Empty} for reuse. \textit{Remove} can also optionally compress probe chains by moving entries further away closer to the original bucket, optimising cache usage. A basic code walk-through follows, which outlines the most important parts of the algorithm.

\begin{figure}
\begin{minipage}{0.01\textwidth}
\hspace{0.05cm}
\end{minipage}
\begin{minipage}{0.44\textwidth}
\begin{programf}
fn Add(key: K) -> bool \{
  key\_hash = hash(key);
  ob = key\_hash \% size;
  rc\_before = table[ob].rc;
  // Part 1: Run an optional read of the table...
  ...
  // Part 2: Reserve a bucket
  rb = ob;
  offset = 0;
  for(; offset < MAX\_DISTANCE; rb++, offset++) \{
  retry:
    v, s = table[rb].vs;
    if(s == Empty) \{
      if(CAS(\&table[rb].vs, 
        \{v, Empty\}, \{v + 1 , Busy\})) \{
        break;
      \} else \{ goto retry; \}
    \}
  \}
  // Part 3: Is the reserved bucket within -
  // general range?
  if(offset < MAX\_DISTANCE) \{ 
    // Is the reserved bucket within -
    // neighbourhood range?
    before\_rb = rb;
    while(offset >= NEIGHBOURHOOD\_DISTANCE) \{ 
        rb, offset = FindCloserBucket(rb, offset);
        // No closer bucket found, resize.
        if(rc == before\_rb) \{
          resize();
          ...
        \}
    \}
    table[rb].k = key;
    table[rb].vs = \{v, Inserting\};
  \} else \{
    // We need to resize the table.
    resize();
    ...
  \}
  // Part 4: Modified Purcell-Harris 
  // exclusivity check.
  ...
\}
\end{programf}
\caption{Pseudo-code for \textit{Add}}
\label{fig:Add}
\end{minipage}
\end{figure}

\begin{flushleft}
\textit{A - Contains}
\end{flushleft}
All line references relate to the code in Figure \textbf{\ref{fig:Contains}} The \textit{Contains} method is relatively unchanged from the blocking version. \textit{Contains} need not worry about interleaving \textit{FindCloserBucket} calls interfering with its correctness, as a relocation counter check simply restarts the method if a change is detected. \textit{Contains} calculates the initial starting bucket and loads the current relocation counter on lines \textbf{2 - 4}. Next it loads the current bit-mask for the original bucket, examining all bits set and calculating all indices to examine for key membership (lines \textbf{7 - 11}). Lines \textbf{13 - 16} load the state of the bucket, check whether the bucket is a \texttt{Member}, and subsequently proceed to check the key for a match. Following an unsuccessful search, lines \textbf{23 - 28} reload the relocation counter to check for a change, returning \texttt{false} if there hasn't been one, and running again if there was.

\begin{flushleft}
\textit{B - Add}
\end{flushleft}

All of the following code lines refer to Figure \textbf{\ref{fig:Add}}. \textit{Add} can be broken down into four sections, one optional and the other three necessary. The first and optional section is to check if the key already exists in the table. The second section involves claiming an \texttt{Empty} bucket, the third involves moving that bucket to within neighbourhood range if necessary, while the fourth performs an exclusiveness check once the bucket is within range. Dealing with the first section, \textit{Add} performs a general preamble for hash tables on lines \textbf{2 - 4}, calculating hash values and loading relocation counters. On line \textbf{6} \textit{Add} can run an optional check of the table, to ensure the item being inserted is not already in the table. The code is more or less identical to that of \textit{Contains} so we leave it out here. Next, in the second section (lines \textbf{8 - 17}), the algorithm attempts to reserve an \texttt{Empty} bucket as \texttt{Busy} up to some defined probe limit. This limit is a user provided parameter. It represents the probe distance tolerated before a table resize is necessary. The same limit exists in the original blocking Hopscotch Hashing.

The third section checks whether the user distance has been violated, causing a resize if so (lines \textbf{22} and \textbf{36 - 37}. If the bucket claimed is within the \texttt{MAX\_DISTANCE}, the algorithm checks if the distance is within neighbourhood range on line \textbf{26}, moving the claimed bucket backwards until it is in range. The algorithm moves the bucket closer by calling \textit{FindCloserBucket} on line \textbf{27}, updating the reserved bucket and its offset every iteration until the bucket is within range. If the method \textit{FindCloserBucket} returns and no progress has been made, the table is considered saturated and a call to \textit{resize} is made on line \textbf{30}. Once it is within neighbourhood range, the bucket has its key written and state updated on lines \textbf{34 - 35}. The fourth and final section is a slightly modified Purcell-Harris uniqueness check, which instead searches within a fixed bound probe, looping on a relocation counter in case of concurrent relocation. The last modifications are to check the relocation counter before and after the exit point in the original Purcell-Harris method. The retry mechanism is identical to the likes of \textit{Contains} and \textit{Remove}.

\begin{figure}
\begin{minipage}{0.01\textwidth}
\hspace{0.05cm}
\end{minipage}
\begin{minipage}{0.44\textwidth}
\begin{programf}
fn Remove(key: K) -> bool \{
  key\_hash = hash(key);
  ob = key\_hash \% size;
  rc\_before = table[ob].rc;
  while(true) \{
    // Load the neighbourhood bit-mask
    bm = table[ob].bm;
    while(bm != 0) \{
      // Find lowest set bit
      lsb = lowest\_set\_bit(bm);
      index = ob + lsb;
    retry:
      // Purcell Harris bucket check.
      \_, state = table[index].vs;
      if(state == Member) \{
        if(table[index].k == key) \{
          if(CAS(\&table[index].vs, 
                 \{v, Member\}, \{v, Busy\})) \{
            // Optionally: shift entries to -
            // closer bucket
            ...
            // Remove key and the bit, -
            // mark bucket as Empty
            table[index].key = Nil;
            table[ob].bm.fetch\_xor(1 << lsb);
            table[index].vs = \{v + 1, Empty\};
            return true;
          \} else \{ goto retry; \}
        \}
      \}
      // Remove the bit just checked.
      bm = XOR(bm, 1 << lsb);
    \}
    // Check the relocation counter
    rc\_after = table[ob].rc;
    if(rc\_before == rc\_after) \{
      return false;
    \}
    // Check bit-mask again
    rc\_before = rc\_after;
  \}
\}
\end{programf}
\caption{Pseudo-code for \textit{Remove}}
\label{fig:Remove}
\end{minipage}
\end{figure}

\begin{flushleft}
\textit{C - Remove}
\end{flushleft}

All of the following code lines refer to Figure \textbf{\ref{fig:Remove}}. \textit{Remove} is near identical to \textit{Contains}, except that when a key match happens, the method tries to \textit{CAS} the state variable from \texttt{Member} to \texttt{Busy} (lines \textbf{17 - 18}). Line \textbf{21} is where one could optionally compress the other entries backward, closer to their original bucket. Lines \textbf{24 - 26} remove the key from the table, remove the bit from the bit-mask, and set the bucket back to \texttt{Empty} for reuse. The algorithm for the preamble on lines \textbf{2 - 4} and relocation counters on lines \textbf{35 - 40} is the same as seen in \textit{Contains}.

\begin{figure}
\begin{minipage}{0.01\textwidth}
\hspace{0.05cm}
\end{minipage}
\begin{minipage}{0.48\textwidth}
\begin{programf}
fn FindCloserBucket(rb: u64, offset: 64) -> \{u64, u64\} \{
  rv, rs = table[rb].vs;
  while(true) \{
  begin:
    // Move back as far as possible
    dist = NEIGHBOURHOOD\_DISTANCE - 1;
    desc = create\_descriptor();
    for(cb = rb - dist; cb < rb; cb++, dist--) \{
      rc\_before = table[cb].rc;
      bm = table[cb].bm;
      while(bm != 0) \{
        lsb = lowest\_set\_bit(bm);
        i = ob + lsb;
        // Check bucket only if advantageous to move
        if(i >= rb) \{ break; \}
        iv, is = table[i].vs;
        // Is this bucket a candidate?
        if(is == Member) \{
          table[rb].k = table[i].k;
          // Mark our bucket as active
          table[cb].bm.fetch\_or(1 << dist);
          // Prepare the K-CAS descriptor
          desc.add(\&table[cb].rc, rc\_before, 
            rc\_before + 1);
          desc.add(\&table[i].vs, \{iv, is\}, 
            \{iv, Busy\});
          desc.add(\&table[rb].vs, \{rv, rs\},  
            \{rv, Member\});
          if(!K\_CAS(desc)) \{
            // Turn off our bit preemptively turned on
            table[cb].bm.fetch\_xor(1 << dist);
            goto begin;
          \}
          // Unmark the now moved bucket, continue on.
          table[cb].bm.fetch\_xor(1 << lsb);
          return \{ i, offset - dist \};
        \}
        // Remove the bit just checked.
        bm = XOR(bm, 1 << lsb);
      \}
      // Check the relocation counter
      rc\_after = table[ob].rc;
      if(rc\_before != rc\_after) \{
        goto begin;
      \}
    \}
    // Return the same bucket and offset to indicate failure.
    return \{ rb, offset \};
  \}
\}
\end{programf}
\caption{Pseudo-code for \textit{FindCloserBucket}}
\label{fig:Closer}
\end{minipage}
\end{figure}

\begin{flushleft}
\textit{D - Find Closer Bucket}
\end{flushleft}

All of the following code lines refer to Figure \textbf{\ref{fig:Closer}}. The goal of \textit{FindCloserBucket} (\textit{FCB}) is to move back some bucket marked as \texttt{Busy} with another bucket which is already a member of the table. The use-case is when the \texttt{Add} method claims a bucket inside the \texttt{MAX\_DISTANCE}, but outside the \texttt{NEIGHBOURHOOD\_DISTANCE}. In this case, displacing a  bucket already in the table is necessary. Like other methods, \textit{FCB} loops while there is a relocation counter discrepancy, since another \textit{FCB} has run and potentially disrupted the result. Lines \textbf{6 - 7} calculate the maximum distance a bucket can be moved, and create a new \textit{K-CAS} descriptor. The loop on line \textbf{8} examines each bucket in an earlier position than the one being relocated. The loop on line \textbf{11} examines the current bucket's bit-mask for a candidate to swap, and checks if that move is worthwhile. Swapping is conditional on two criteria. The first is that moving the candidate entry keeps it within its neighbourhood, the second is that the entry being swapped actually moves closer to its home neighbourhood. The first criterion is ensured by definition, as the loop's variables are initialised to all be legal swaps (lines \textbf{6 - 8}). The second criterion is checked on line \textbf{15} to ensure the swap actually moves the bucket closer to home.

Once a candidate bucket has been identified, its bit-mask is marked to include the destination bucket (line \textbf{21}). Lines \textbf{23} - \textbf{28} fill out the descriptor with all the information needed to swap the two buckets and increment the relocation counter. The \textit{K-CAS} function is invoked on line \textbf{27}, either atomically swapping the two buckets, or else failing. Failing to swap the buckets results in the candidate's bit-mask having the destination bit unmarked, and the method is restarted. If the call succeeds, then the bit where the candidate used to be is unmarked from its bit-mask and the new bucket is returned on lines \textbf{35} and \textbf{36}. Lines \textbf{42 - 44} check for a relocation counter discrepancy, restarting the method if one is detected.

\section{Proof Of Correctness and Progress}

\subsection{Correctness}
We present a simple sketch proof of correctness using lemmas to build up our proof argument. Each method will be evaluated for \textit{linearisability} \cite{Herlihy:1990:LCC:78969.78972}. If every method is \textit{linearisable}, then the entire object is \textit{linearisable}. We deal with every code point in every method, highlighting the particular \textit{linearisation} point in the algorithm and in the code.

The sketch of our proof argument is as follows. Both \textit{Contains} and \textit{Removes} read the relocation counters, then the bit-mask neighbourhood, and perform a combination of Hopscotch and Purcell-Harris state-based reads. After an operation is performed, the relocation counter is re-read, and the operation is performed again if a relocation is detected. Both methods can therefore be considered in isolation, as any abnormalities caused by moving entries around the table are dealt with by the relocation counters. A \textit{linearisable} \textit{Add} method must not fail to insert a key where there isn't one, and must not succeed in inserting a key where there already is one. The only part which makes a key a member of the table is the uniqueness check. This last component of \textit{Add}, the uniqueness check, has to deal with concurrent \textit{Add} calls moving entries around. It is the only component capable of spuriously making a key a \texttt{Member}, or deleting it. The first section is a linear probe to claim an \texttt{Empty} bucket as \texttt{Busy}, claiming it via a \textit{CAS}. Once a bucket is in the \texttt{Busy} state, it can only transition to another state by the thread that marked it as \texttt{Busy}. In other words, the thread has pseudo ownership of the bucket. The linear probing algorithm is simple and doesn't create any difficulty in reasoning about the concurrent correctness of the algorithm. All that matters is that a bucket is moved from \texttt{Empty} to \texttt{Busy}. The second stage is to move that claimed bucket to within neighbourhood range if not already inside. Moving is achieved by linearly probing towards the claimed bucket and atomically swapping it with a valid bucket found along the way. The probing for another bucket is performed from the max distance the bucket could move, that is, from a ``neighbourhood distance'' away. The atomic swap is accomplished by \textit{K-CAS} so that swapping has no visible intermediate state and can retry if the operation fails. Swapping increments the relocation counters for the home bucket, forcing both \textit{Contains} and \textit{Remove} to re-run if necessary. Once the entry is moved within neighbourhood range, a uniqueness check is performed. The check is near identical to that found in the Purcell-Harris table, the only difference being a single extra step. The method must check the relocation counter before attempting to commit an entry to a \texttt{Member} state, as a relocation could lead to an incorrect result (relocating the entry already in a \texttt{Member} state and thus missed by the uniqueness check) and so the method would need to be restarted as per \textit{Contains} and \textit{Remove}.

\begin{flushleft}
\textbf{Lemma 1:} \textit{Contains} is \textit{Linearisable}.
\end{flushleft}

\begin{flushleft}
Proof: \textit{Contains} initially loads the relocation count on line \textbf{4}, creating a basic snapshot of the bucket. The bucket's bit-mask is loaded on line \textbf{7}, and each entry is checked. \textit{Contains} loads the key on line \textbf{15}, and is the \textit{linearisation} point for a successful \textit{Contains} call. If a matching key is not found, then the relocation counter is checked again on line \textbf{23} to ensure a matching key hasn't been moved around during the search. This re-load indicates whether the snapshot was invalidated during the search, and is the \textit{linearisation} point for an unsuccessful \textit{Contains} call. All code paths in \textit{Contains} have \textit{linearisation} points and thus \textit{Contains} is \textit{linearisable}.
\end{flushleft}

\begin{flushleft}
\textbf{Lemma 2:} \textit{Add} is \textit{Linearisable}.
\end{flushleft}

\begin{flushleft}
Proof: \textit{Add} is composed of three primary parts and one optional part. The first optional operation can cut \textit{Add} short by determining that a key is already present in the table, and returning \texttt{false}. The optional check has the same \textit{linearisation} points at \textit{Contains}. Once a bucket is marked as \texttt{Busy} on line \textbf{13 - 14}, the algorithm moves the bucket into the appropriate range with repeated calls to \textit{FindCloserBucket}. If \textit{FindCloserBucket} fails to find a bucket, then the table is considered saturated and a resize commences. Once the bucket is within range, the key is written into the bucket and the state is changed to \texttt{Inserting}. Once the bucket has transitioned to \texttt{Inserting}, then a modified Purcell-Harris exclusivity check is run. There is only one modification point, that being to check the bucket relocation counter before attempting to mark a key as a \texttt{Member}. However, this modification doesn't change the \textit{linearisation} point, just whether the method retries. All code paths in \textit{Add} have \textit{linearisation} points and thus \textit{Add} is \textit{linearisable}.
\end{flushleft}

\begin{flushleft}
\textbf{Lemma 3:} \textit{Remove} is \textit{Linearisable}.
\end{flushleft}

\begin{flushleft}
Proof: \textit{Remove} is similar to \textit{Contains}, except a \textit{CAS} is attempted on the bucket state to move it from \texttt{Member} to \texttt{Busy}. The \textit{CAS} on line \textbf{17} represents the \textit{linearisation} point for a successful remove. \textit{Remove} has the same \textit{linearisation} points as \textit{Contains} when searching for an entry. All code paths in \textit{Remove} have \textit{linearisation} points and thus \textit{Remove} is \textit{linearisable}.
\end{flushleft}

\begin{flushleft}
\textbf{Lemma 4:} \textit{FindCloserBucket} is \textit{Linearisable}.
\end{flushleft}

\begin{flushleft}
Proof: \textit{FindCloserBucket} attempts to atomically swap the current bucket in a state of \texttt{Busy} with another bucket already marked as a \texttt{Member}. The method begins by linearly probing from a certain distance away to find a candidate bucket. The bit-mask is checked for possible buckets; buckets which move the bucket being relocated further from its home are excluded (lines \textbf{11 - 15}). Once a candidate has been identified, the bucket being relocated is added to the candidate bucket's bit-mask (line \textbf{21}) in anticipation of the bucket swap. The swapping of the buckets is executed on line \textbf{29} by \textit{K-CAS}. A failed \textit{K-CAS} means that the bucket has either been deleted or moved by a concurrent call, requiring the preemptively set bit to be unset and the method restarted. If the \textit{K-CAS} succeeds, then this is the \textit{linearisation} point of a successful call to \textit{FindCloserBucket}. The method must also remove the old location just moved from the bit-mask, returning the new offsets. If no candidate buckets are found during the linear probe, then the method returns the old offsets. The \textit{linearisation} point for a failed call is the last check of the relocation counters on line \textbf{43}. All code paths in \textit{FindCloserBucket} have \textit{linearisation} points and thus \textit{FindCloserBucket} is \textit{linearisable}.
\end{flushleft}

\begin{flushleft}
\textbf{Theorem 1:} The hash table is \textit{Linearisable}.
\end{flushleft}

\begin{flushleft}
Proof: Each method of the hash table is \textit{linearisable} as per \textit{Lemma 1}, \textit{2}, \textit{3}, and \textit{4}. Hence the hash table is \textit{linearisable}.
\end{flushleft}

\subsection{Progress}
Progress, like correctness, will be argued informally, as the base table of Purcell-Harris already has strong progress arguments accompanying its publication. Both \textit{Contains} and \textit{Remove} methods can be made re-run if run concurrently with a \textit{FindCloserBucket} call relocating an entry from its relevant neighbourhood. This re-run, however, implies the success of another call, thus achieving system progress and lock-freedom. \textit{Remove} tries to \textit{CAS} the state variable into \texttt{Busy} from \texttt{Member}, potentially restarting if the \textit{CAS} fails. Failure here means the success of another \textit{Remove} or a relocation in \textit{FindCloserBucket}; either way, progress has been achieved. \textit{FindCloserBucket} has the same behaviour as \textit{Contains} and \textit{Remove}, that is, re-running if any relocation counters have been changed since the initial snapshot. \textit{FindCloserBucket} also re-runs if the call to \textit{K-CAS} fails. Such failure only occurs if another method changed the state variable (\textit{Remove} or another \textit{FindCloserBucket}), meaning some other process made progress. The main components of the \textit{Add} method are also lock-free. A simple linear probe with a \textit{CAS} loop will contend with other linear probes, but the failure of one means the success of another, meeting the standards of lock-freedom. Finally, as per standard Purcell-Harris, the uniqueness check is lock-free. Our extra step of checking the relocation counter forces the method to run again, which implies that some other method has made progress on the object, again ensuring lock-freedom. On the whole, we argue that since all methods are lock-free, then the object as a whole must be lock-free.

\section{Performance and Discussion}
In this section we detail the performance and implementation of our algorithm. All of our code is made freely available online \cite{LockFreeHopscotchBenchmark}. This includes lock-free Hopscotch Hashing, implementations of alternative competing algorithms (either coded by us or obtained via online sources), and microbenchmarking code which allows readers to replicate our results.

\subsection{Experimental Setup}
For our experiments we opted to use a set of microbenchmarks stressing the hash-table under various capacities and workloads. Our benchmarks were run on a 4 CPU machine, with each CPU (Intel(R) Xeon(R) CPU E7-8890 v3) featuring 18 cores with two hardware threads, and a total of 512 GiB RAM. The machine ran Ubuntu 14.04 with a Linux Kernel version of 3.13.0-141. Each thread was pinned to a specific core for the duration of the test, and threads were scaled in increments of 9, from 9 to 144 threads. When scaling the number of threads, care was taken to pin the thread to an unused core instead of exercising \textit{HyperThreading\texttrademark}. We avoided the use of \textit{HyperThreading\texttrademark} for as long as possible, scheduling threads to another \textit{NUMA} CPU to avoid its use. \textit{HyperThreading\texttrademark} was only used after every core on each CPU had one thread pinned to it \textit{HyperThreading\texttrademark}.

\begin{figure}[!htbp]
    \centering
    \includegraphics[width=0.40\textwidth]{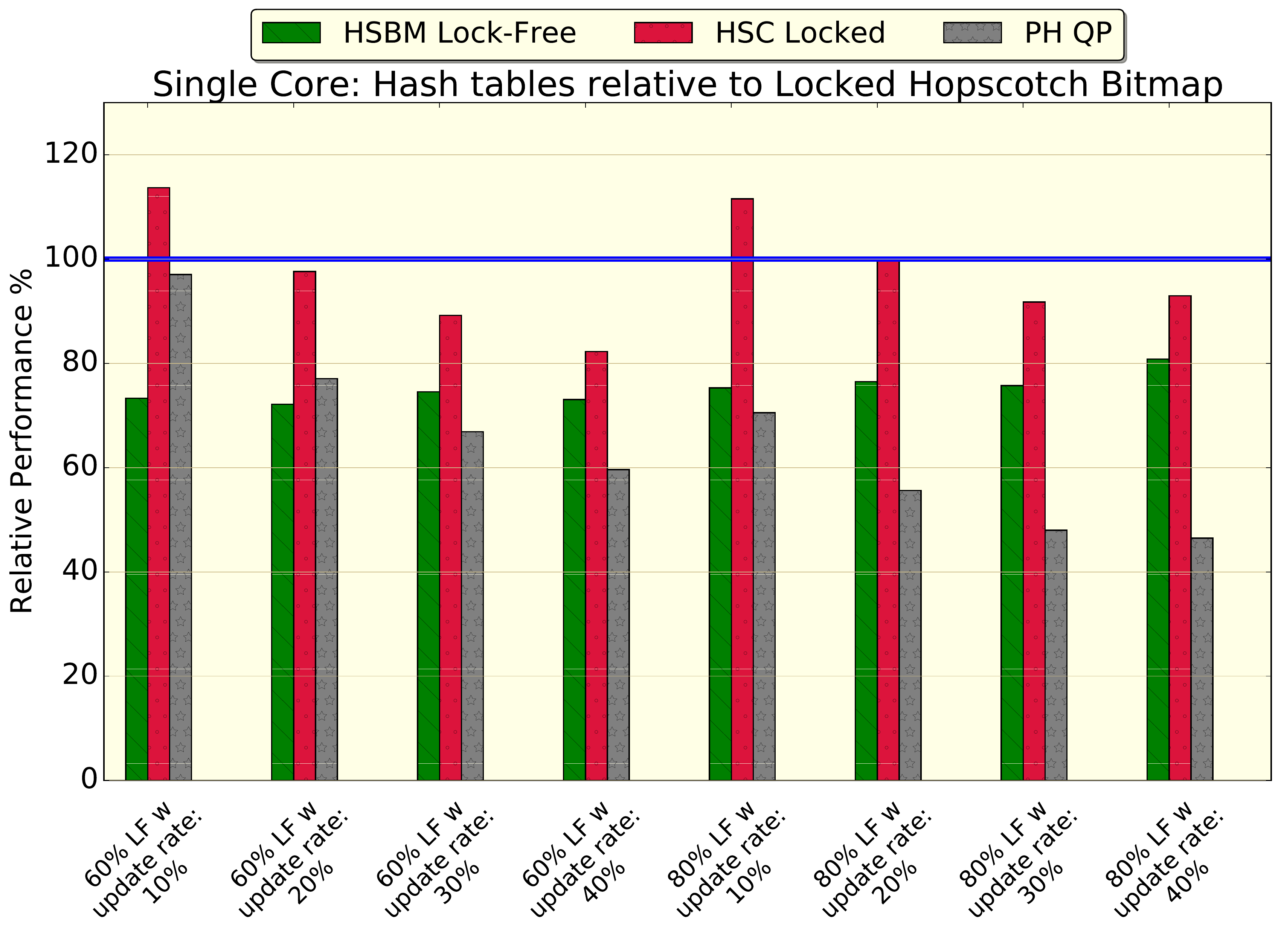}
    \caption{Single thread performance relative to Locked Hopscotch Bit-Map.}
    \label{fig:SingleThreadMit}
\end{figure}

\begin{figure*}[!htbp]
    \centering
    \begin{subfigure}[b]{0.38\textwidth}
        \includegraphics[width=\textwidth]{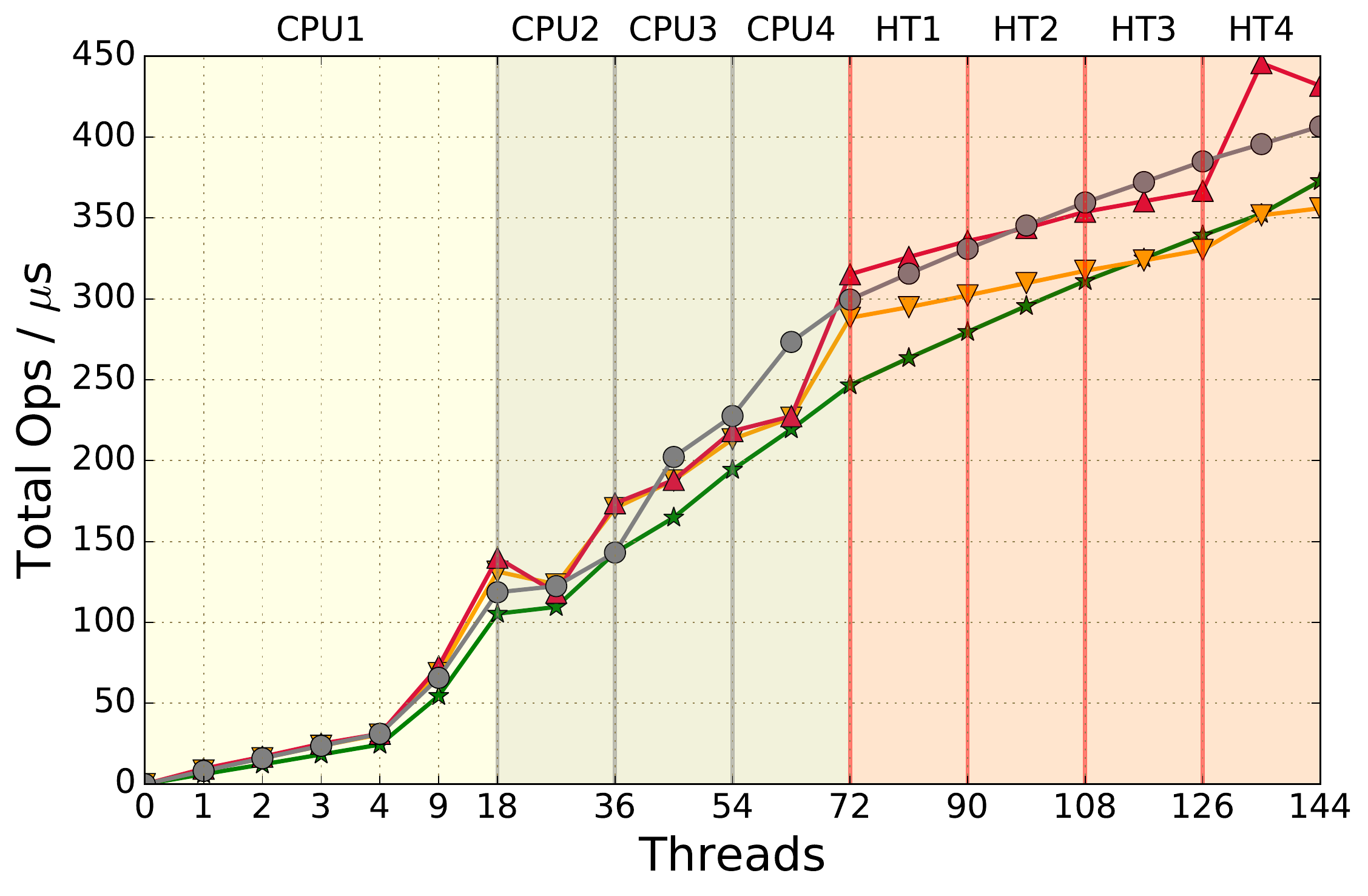}
        \caption{60\% load factor @ 10\%}
        \label{fig:PerfTen60Mit}
    \end{subfigure}
    \qquad
    \begin{subfigure}[b]{0.38\textwidth}
        \includegraphics[width=\textwidth]{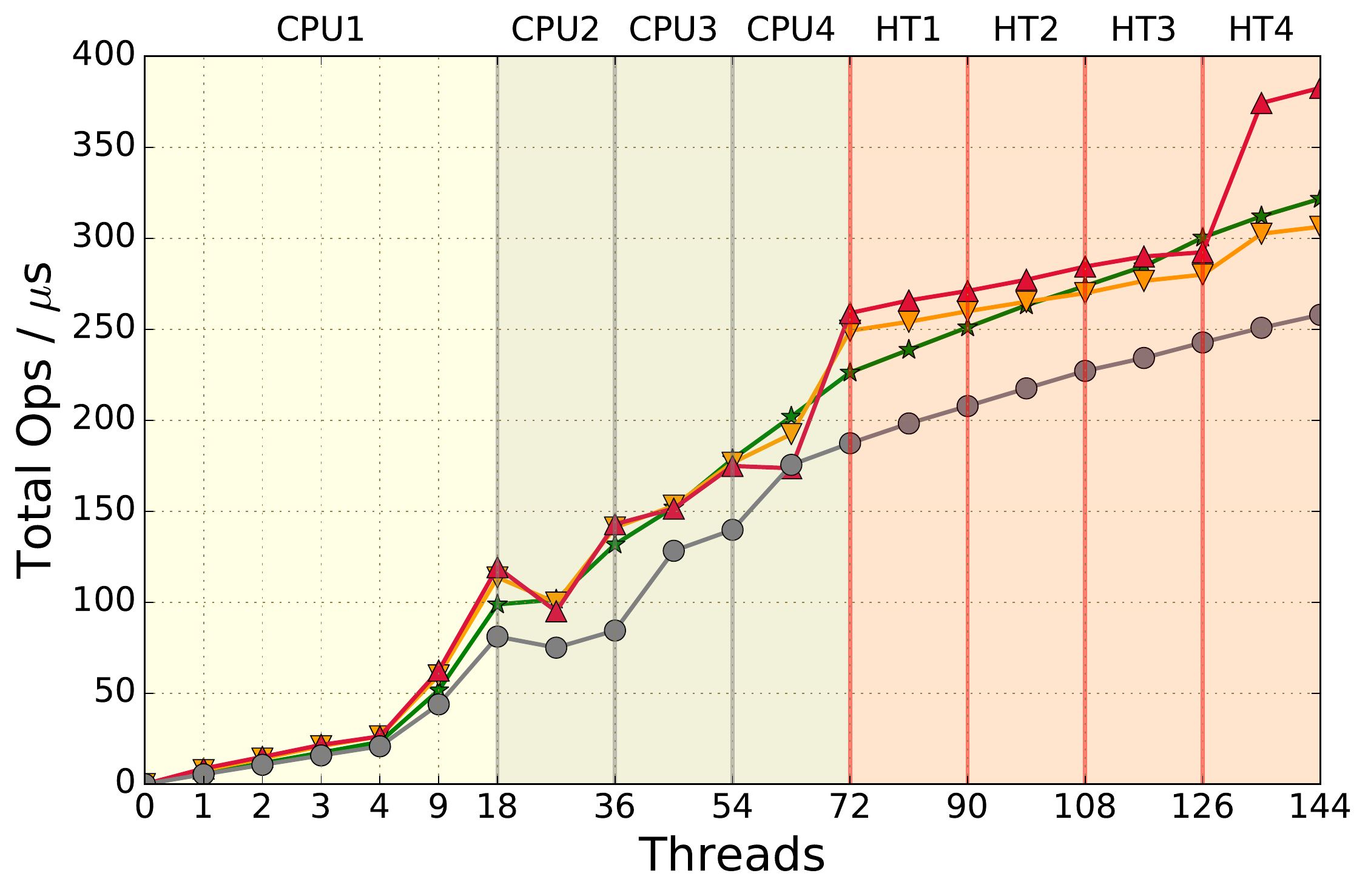}
        \caption{80\% load factor @ 10\%}
        \label{fig:PerfTen80Mit}
    \end{subfigure}
  
    \begin{subfigure}[b]{0.38\textwidth}
        \includegraphics[width=\textwidth]{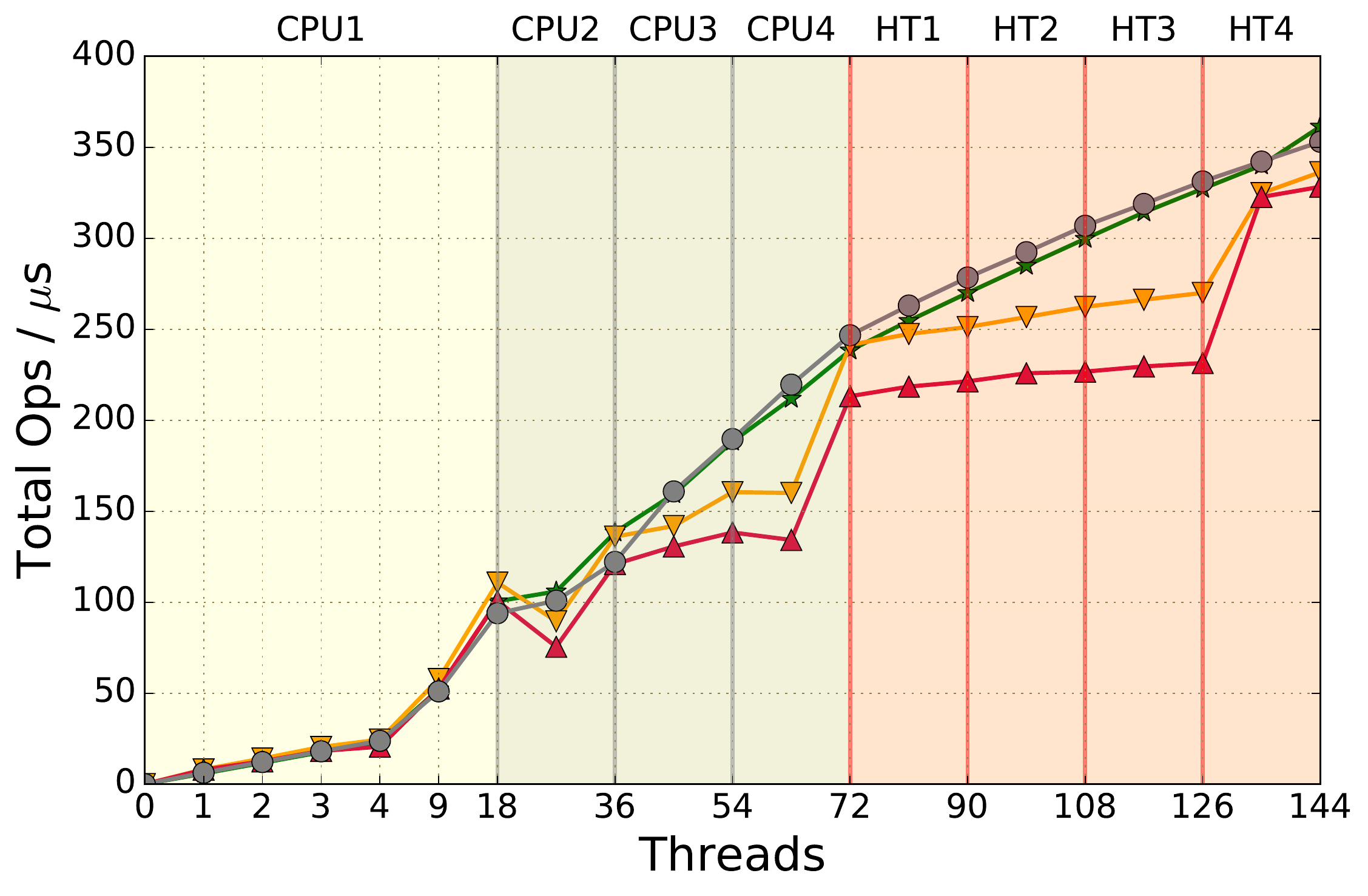}
        \caption{60\% load factor @ 20\%}
        \label{fig:PerfTwenty60Mit}
    \end{subfigure}
    \qquad
    \begin{subfigure}[b]{0.38\textwidth}
        \includegraphics[width=\textwidth]{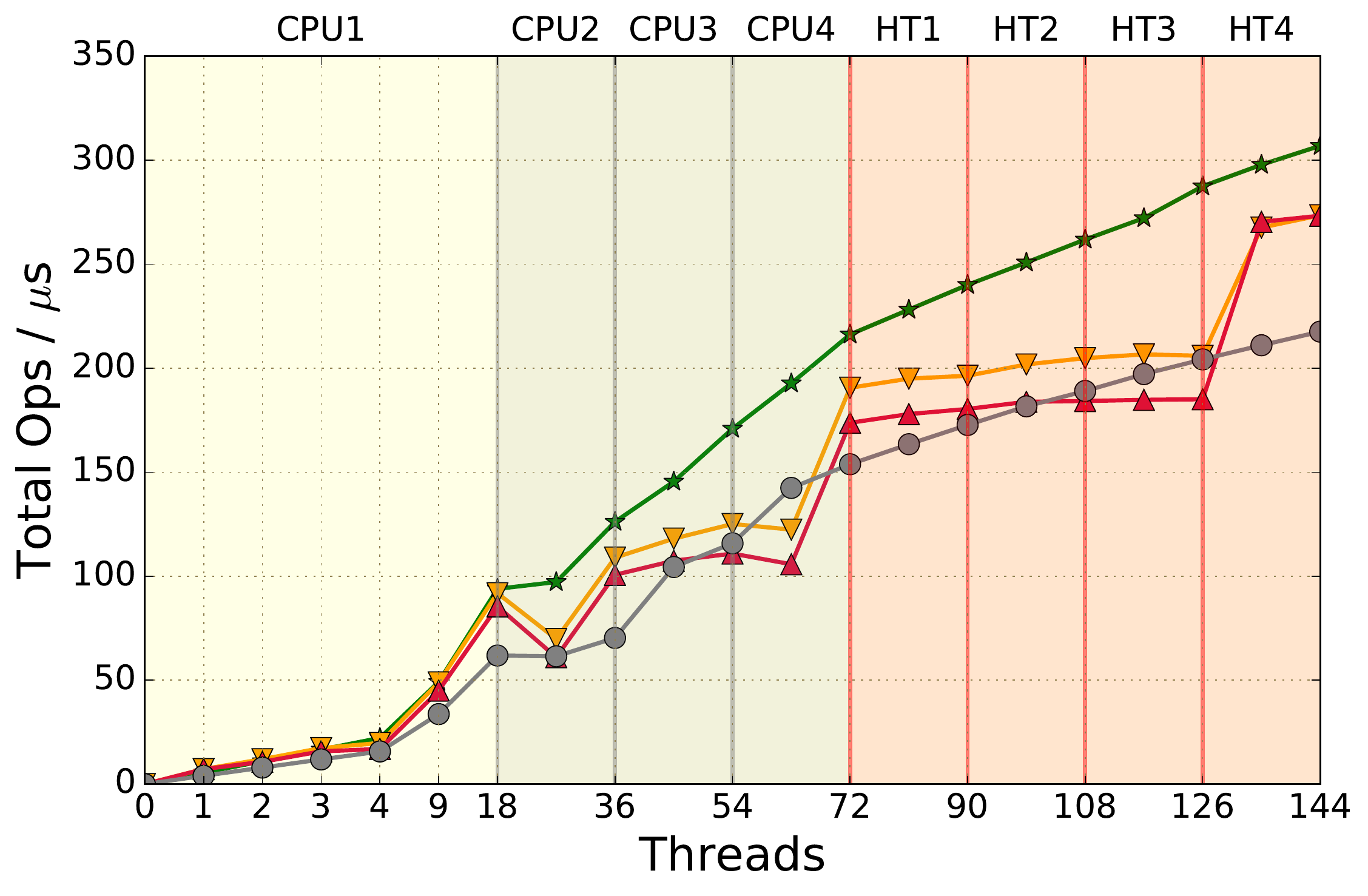}
        \caption{80\% load factor @ 20\%}
        \label{fig:PerfTwenty80Mit}
    \end{subfigure}
    
    \begin{subfigure}[b]{0.8\textwidth}
        \includegraphics[width=\textwidth]{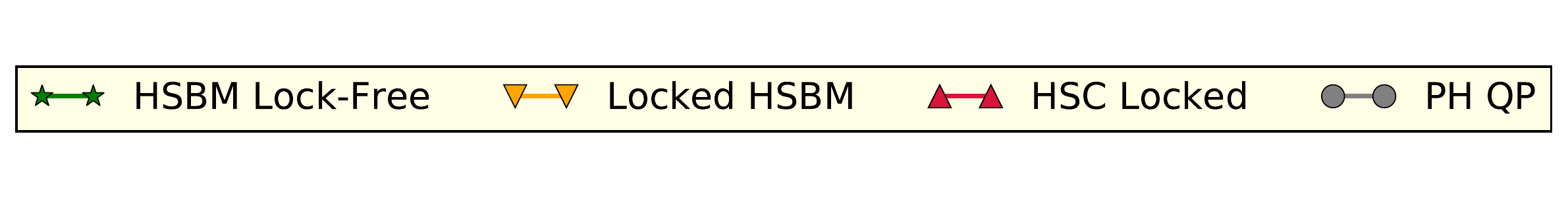}
    \end{subfigure}
    \caption{Performance graphs for low update rate.}
    \label{fig:PerfMitLower}
\end{figure*}


The hash-table algorithms benchmarked include both blocking Hopscotch Hashing with fixed-size bit-masks (HSBM Locked), the relative offset variant with probe-chain compression (HSC Locked) \cite{Hopper}, the Purcell-Harris Quadratic Probing (PH QP) hash-table \cite{Purcell2005}, and our lock-free Hopscotch Hashing (HSBM Lock-Free).

\begin{figure*}[!htbp]
    \centering
    \begin{subfigure}[b]{0.38\textwidth}
        \includegraphics[width=\textwidth]{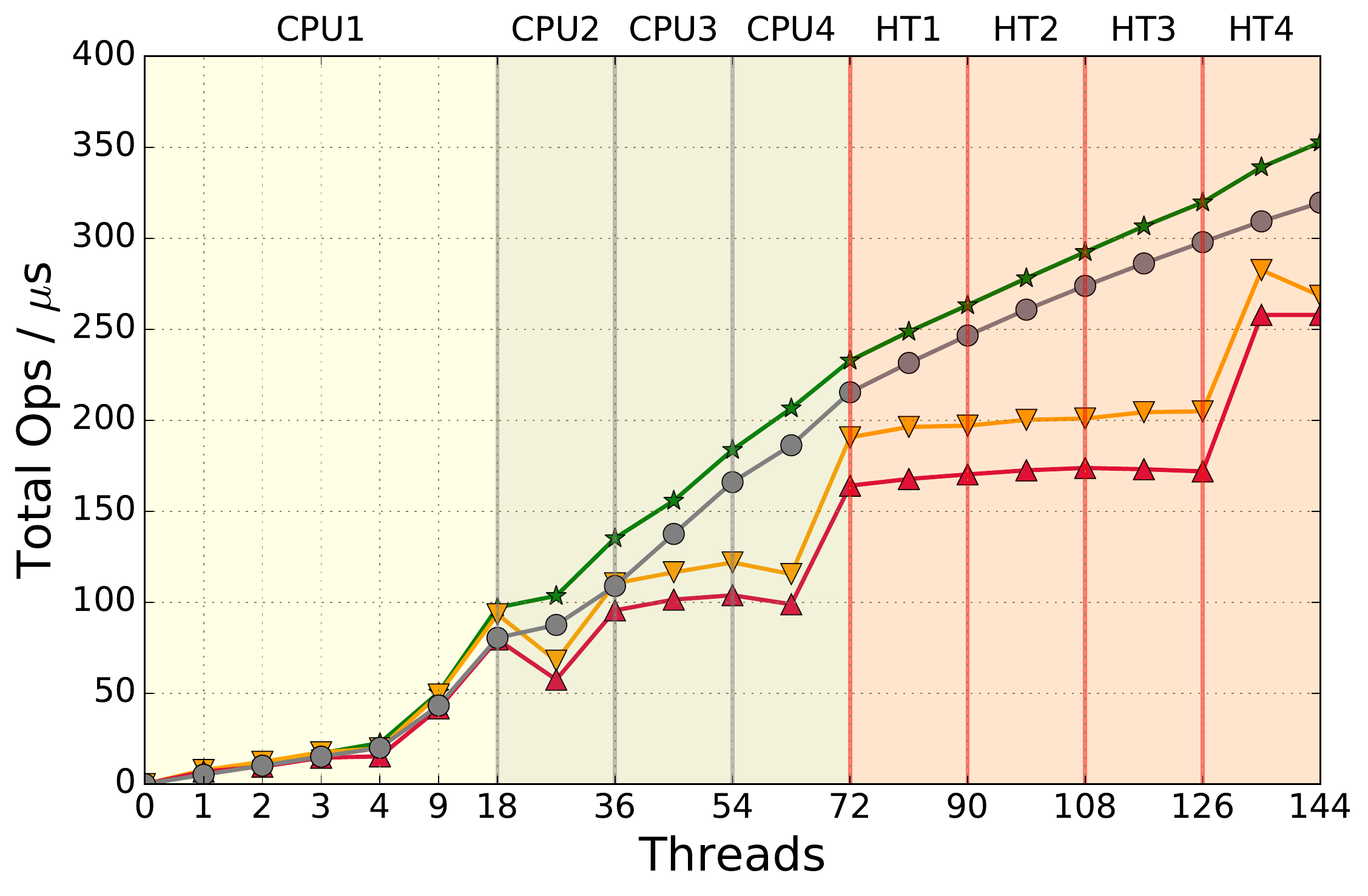}
        \caption{60\% load factor @ 30\%}
        \label{fig:PerfThirty60Mit}
    \end{subfigure}
    \qquad
    \begin{subfigure}[b]{0.38\textwidth}
        \includegraphics[width=\textwidth]{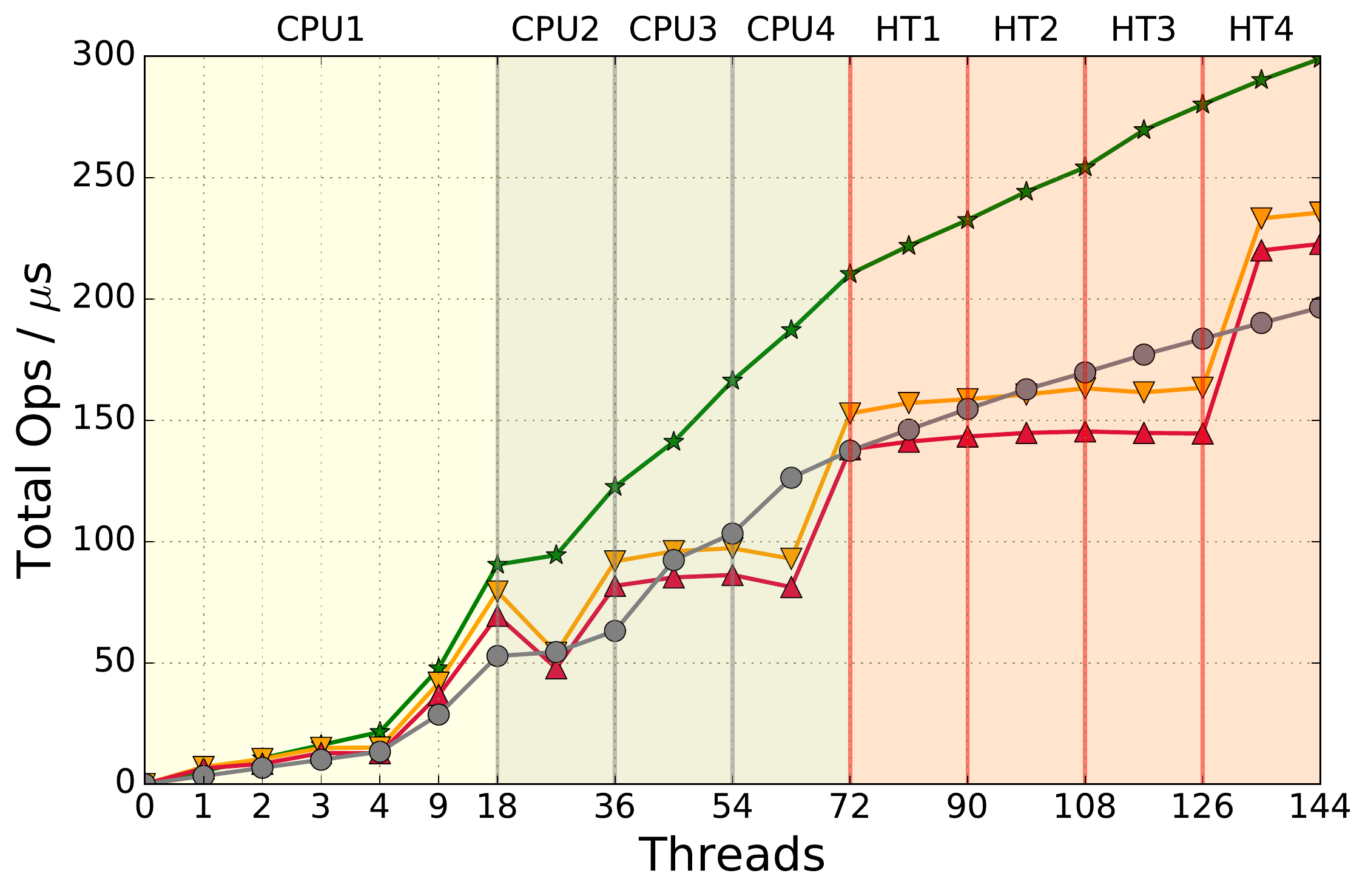}
        \caption{80\% load factor @ 30\%}
        \label{fig:PerfThirty80Mit}
    \end{subfigure}
    
    \begin{subfigure}[b]{0.38\textwidth}
        \includegraphics[width=\textwidth]{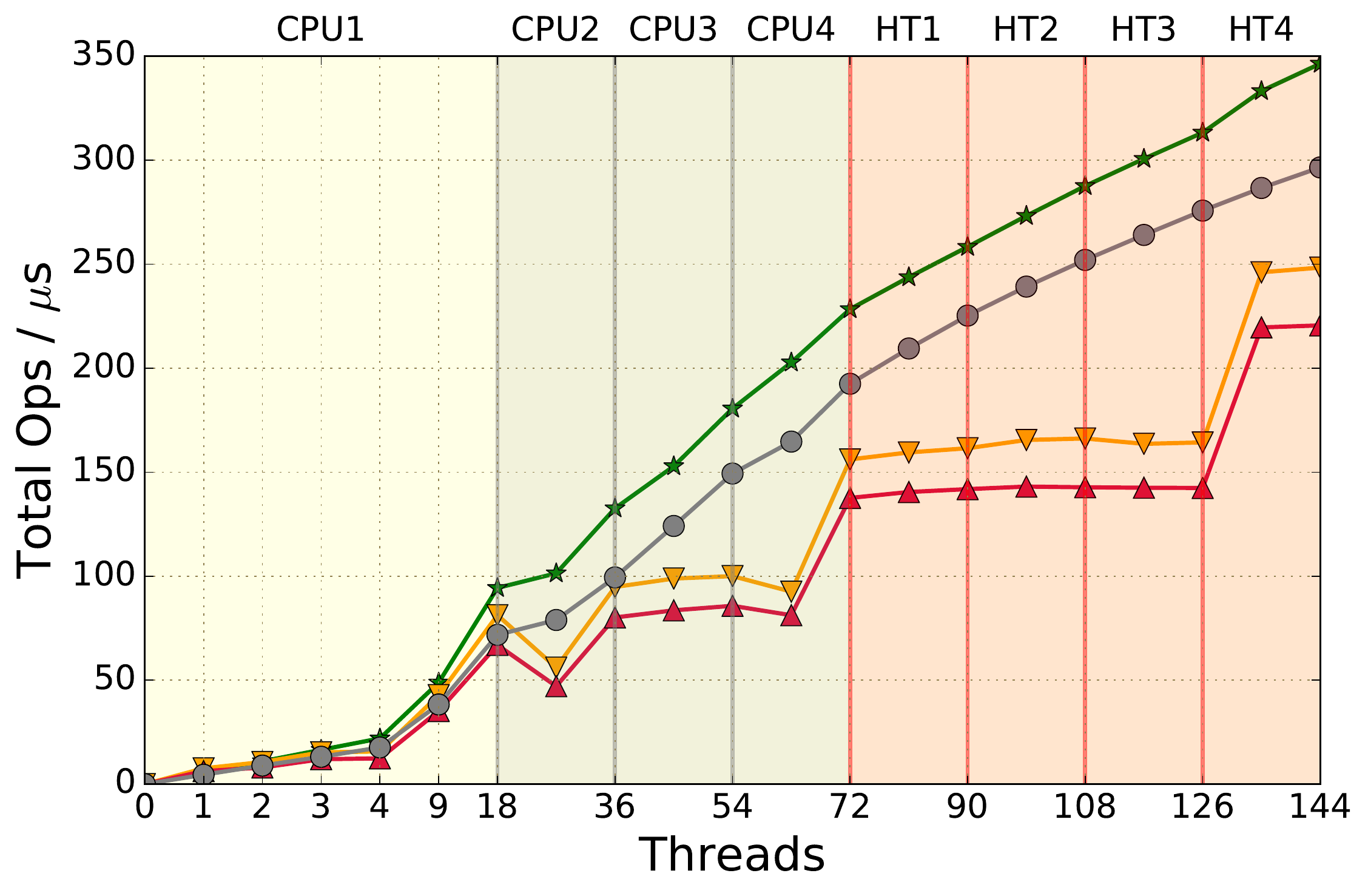}
        \caption{60\% load factor @ 40\%}
        \label{fig:PerfFourty60Mit}
    \end{subfigure}
    \qquad
    \begin{subfigure}[b]{0.38\textwidth}
        \includegraphics[width=\textwidth]{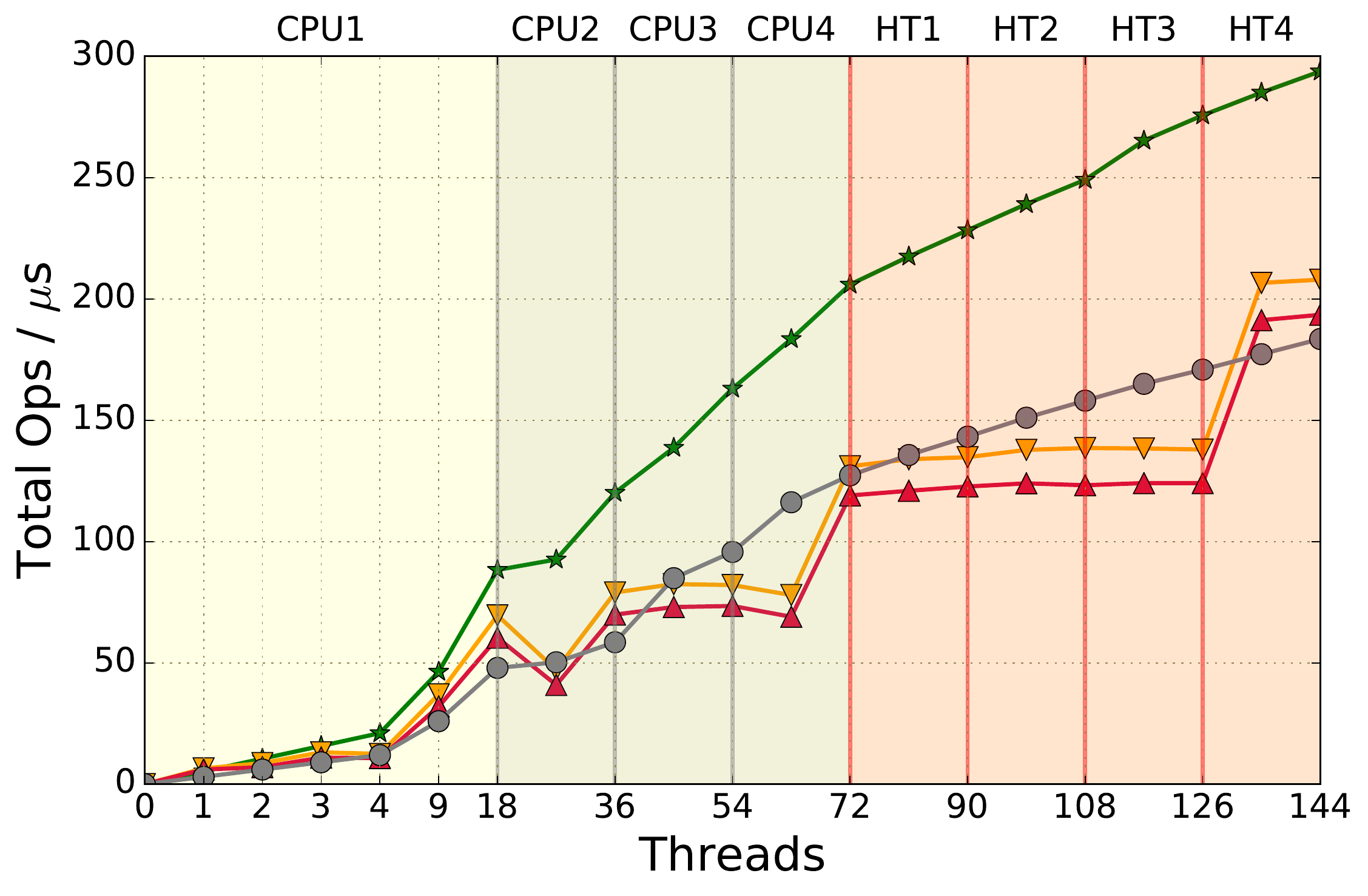}
        \caption{80\% load factor @ 40\%}
        \label{fig:PerfFourty80Mit}
    \end{subfigure}
    
    \begin{subfigure}[b]{0.8\textwidth}
        \includegraphics[width=\textwidth]{graphs/Hopscotchlegend.pdf}
    \end{subfigure}
    
    \caption{Performance graphs for higher update.}
    \label{fig:PerfMitHigher}
\end{figure*}

A number of workload configurations were used in graphing the results. Two load factors of 60\% and 80\% were chosen, along with four read/write workload configurations, namely 90\% reads to 10\% updates, 80\% reads to 20\% updates, 70\% reads to 30\% updates, and 60\% reads to 40\% updates. Updates consist of balanced insertions and deletions. All workload configurations were benchmarked at the specified load factors. We sized the tables at $2^{25}$, as carried out in \cite{Nielsen:2016:SLH:3016078.2851196}. This meant that the table wouldn't fit into the cache, thus highlighting the effective cache use achieved by each algorithm. No memory reclaimer was used, as none was necessary. We used the \texttt{numactl} command to mitigate any negative \textit{NUMA} memory effects. This command specified that allocation could only be carried out on the RAM banks closest to the running CPUs once they came into use following increased thread counts.

Concurrent benchmarking has been refined over the years. We strive to perform a like for like comparison, hence our load factors, read/write workloads, and testing process are similar to a number of other previous concurrent hash-table publications \cite{kelly_et_al:LIPIcs:2018:10070}, \cite{Nielsen:2016:SLH:3016078.2851196}, \cite{Hopper}. The process was carried out as follows. Each thread called a random method with a random argument from some predefined method and key distribution. All threads were synchronised before execution on the data-structure, and executed for a specified amount of time, rather than a specific number of iterations. Each thread counted the number of operations it performed on the structure during the benchmark. The total amount of operations per microsecond for all threads was then graphed, showing throughput. Each experiment was run five times for 10 seconds each, and the average of each result was computed and plotted. All of our algorithms were written in C++11, and compiled with \textit{g++} 4.9.4. The compiler had \texttt{O3} level of optimisation, and also targeted the specific processor architecture it was being run on.

\subsection{Results}
The results are listed in Figures \textbf{\ref{fig:PerfMitLower}} and \textbf{\ref{fig:PerfMitHigher}}, showing the throughput as a function of concurrent threads. Each graph shows the throughput of each algorithm at a lower thread count of 1 - 4 thread(s), so as to better understand the lower scaling. As the number of threads increases, they cross \textit{NUMA} boundaries at multiples of 18. These boundaries are marked by faint grey lines and a background shading in each figure. Once the number of threads exhausts all physical cores, it begins to use \textit{HyperThreading\texttrademark} at 72+ threads. The graphs have another faint red line and background shading to indicate when a new CPU activates \textit{HyperThreading\texttrademark}.

We attempt to identify common trends in all graphs before addressing the specifics of each benchmark result. The common pattern is that, at low thread counts, all algorithms are very competitive in terms of performance. The algorithms typically stay close together in performance up until the number of threads scheduled requires the use of another CPU (18 $<$ threads). There are several ``kinks'' in the graph which are common throughout. All algorithms suffer performance penalties when using an extra CPU (18 $<$ threads, 36 $<$ threads, and 54 $<$ threads). The penalty is either a dip in overall performance or a reduction in the slope of the performance line. As the number of threads increases, all algorithms demonstrate continued slowing in performance, with the angle of the line decreasing further when \textit{HyperThreading\texttrademark} is engaged (72+ threads). The reader should also note that the throughput scaling in each graph is different. In Figure \textbf{\ref{fig:SingleThreadMit}} we highlight the performance of each algorithm using only 1 thread relative to the locked bit-map Hopscotch implementation. The results show that the lock-free algorithms perform significantly worse when using a single thread. This would hint that the majority of ``work'' being done is for scalability when more threads are scheduled.

Another common trend is that, as load factor increases, the performance of a table decreases. As the table fills up with entries, the cost of each operation also increases. The number of collisions goes up, more entries need to be checked, and generally more work is done. The Purcell-Harris table is hit particularly hard by an increase in load factor. The quadratic probing algorithm needs to check substantially more entries than the equivalent Hopscotch tables, leading to a severe drop in performance as load factor increases. As the update rate climbs, the performances of all tables drop, matching the typical expectation for concurrent objects. The locking Hopscotch tables are hit the hardest from the increase in update ratio, doing best under the lightest updates, with lock-free Hopscotch doing the worst. The reason the lock-free algorithms fare worse at light update rates is that they do more work that allows for greater scalability under heavier load. Lastly, the reader will notice the large jumps in the performance of both locked variants of Hopscotch Hashing. The reason here is that the number of locks is set to the number of active threads, as per the original paper \cite{Hopper}. We, however, apply an optimisation which increases the number to the closet power of two. As a result, thread counts near but just under a power of two suffer a performance drop, while those just after see a large increase: the total amount of concurrency in the hash-table has doubled, while the amount of threads has only increased by 9.

Each performance graph is grouped by the update rate and load factor. Accordingly, we analyse the performance according to that grouping. The following analysis refers to the figures in Figure \textbf{\ref{fig:PerfMitLower}}. A mixed bag of winners is evident, as every table switches out for first place depending on the load factor or update rate. Generally, at lower update rates locked Hopscotch fares well, as seen in Figures \textbf{\ref{fig:PerfTen60Mit}} and \textbf{\ref{fig:PerfTen80Mit}}. As the update rate increases, the lock-free algorithms start to pull ahead. Figure \textbf{\ref{fig:PerfTwenty60Mit}} has the lock-free algorithms drawing for first place throughout most of the graph, only to be bested at the last configuration. Figure \textbf{\ref{fig:PerfTwenty80Mit}} shows quadratic probing falling off, with lock-free Hopscotch growing and maintaining its lead over the locked variants in Figure \textbf{\ref{fig:PerfTwenty80Mit}}. We switch focus now to the Figures in \textbf{\ref{fig:PerfMitHigher}}. Broadly speaking, at a higher update rate the throughputs of all algorithms are reduced. The locked variants of Hopscotch Hashing have a significant performance drop at all load factors and update rates. Similarly, the performance of Lock-Free Hopscotch Hashing decreases as the load factor goes up. The gap between Purcell-Harris Quadratic Probing and Lock-Free Hopscotch Hashing grows, with lock-free Hopscotch strengthening its lead at 60\% and 80\% load factor when the update rate increases. 

As is apparent from Figures \textbf{\ref{fig:PerfMitLower}} and \textbf{\ref{fig:PerfMitHigher}}, lock-free Hopscotch starts slow but ends up dominating in terms of performance. The lock-free Purcell-Harris quadratic-probing has a good showing, but drops significantly in performance at the higher load factors. Although our algorithm is slower than the locked Hopscotch and Purcell-Harris tables at the lower updates rates, as both the load factor and update rate increase, our algorithm pulls ahead of the competition. Locked Hopscotch performs best at low update rates and is very strong at all load factors. It performs consistently at each load factor, though suffers considerably under heavy update rates. Overall, the lock-free Hopscotch solution finishes either roughly 20\% behind locked Hopscotch, or 50\% ahead at the highest thread count.

\section{Conclusion and Future Work}
We have presented a lock-free Hopscotch Hashing algorithm which achieves noteworthy performance relative to other lock-free and concurrent algorithms. To the best of our knowledge, this is the first presentation of such an algorithm in the literature. Our experiments show that the approach is competitive with locked Hopscotch at low updates, and dominates above that. The algorithm is relatively simple and just as portable as competitors, needing only single word compare-and-swap instructions. In future work we plan to create a lock-free relative-offset variant and larger bit-mask to potentially improve performance.

\section*{Acknowledgements.}
We wish to thank Nir Shavit for his advice and use of his computing cluster when generating results. We also wish to thank William M. Leiserson for his feedback and critique of this work, as well as his friendship and support. Finally we would like to thank the Irish Centre for High-End Computing (ICHEC) for the use of their machines for benchmarking.

\appendix
\section{Appendix}

\subsection{Additional performance results}
\begin{figure}[!htbp]
    \centering
    \includegraphics[width=0.40\textwidth]{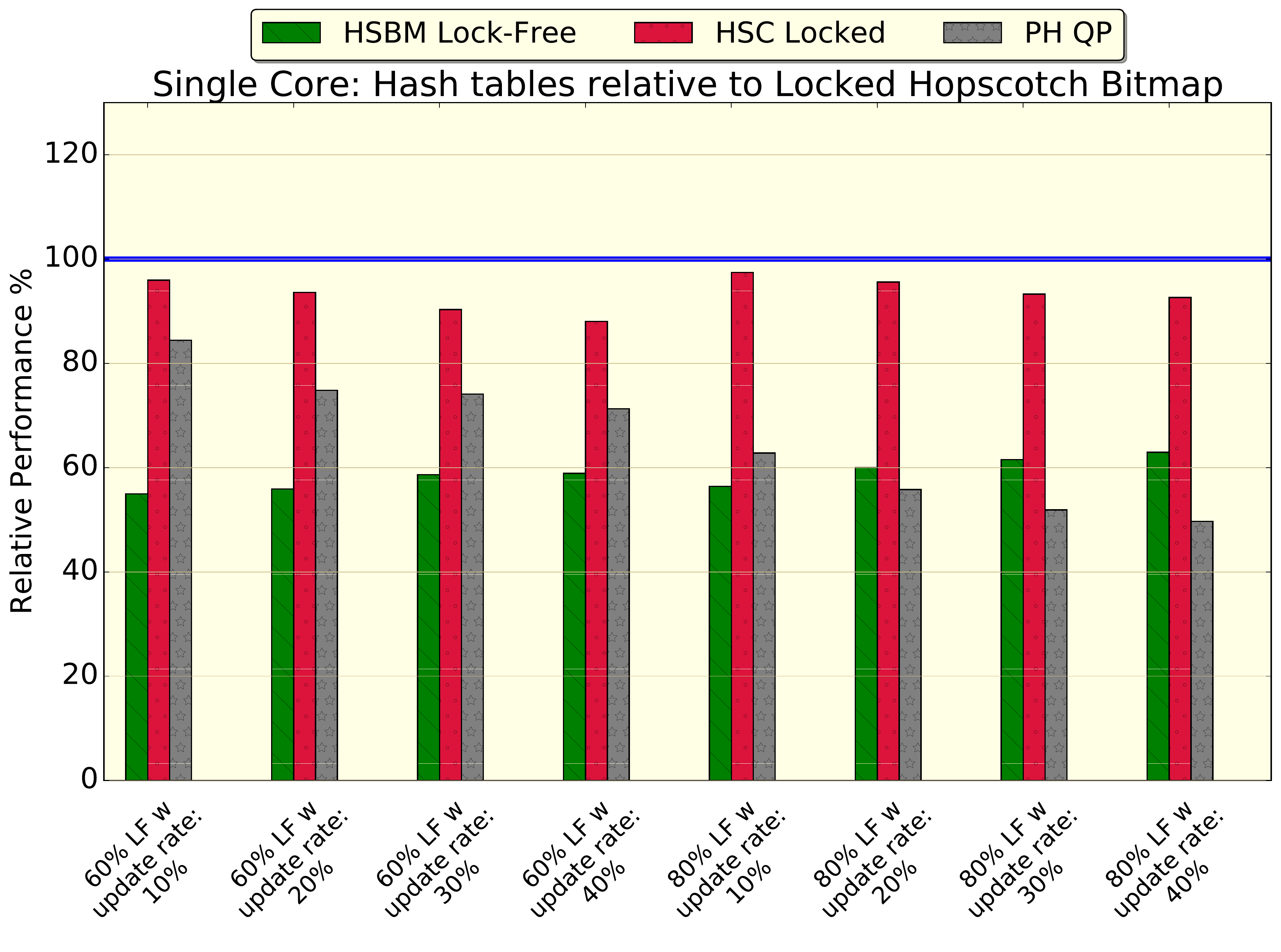}
    \caption{Single thread performance relative to Locked Hopscotch Bit-Map for the second machine.}
    \label{fig:SingleThreadIchec}
\end{figure}

\begin{figure*}[!htbp]
    \centering
    \begin{subfigure}[b]{0.38\textwidth}
        \includegraphics[width=\textwidth]{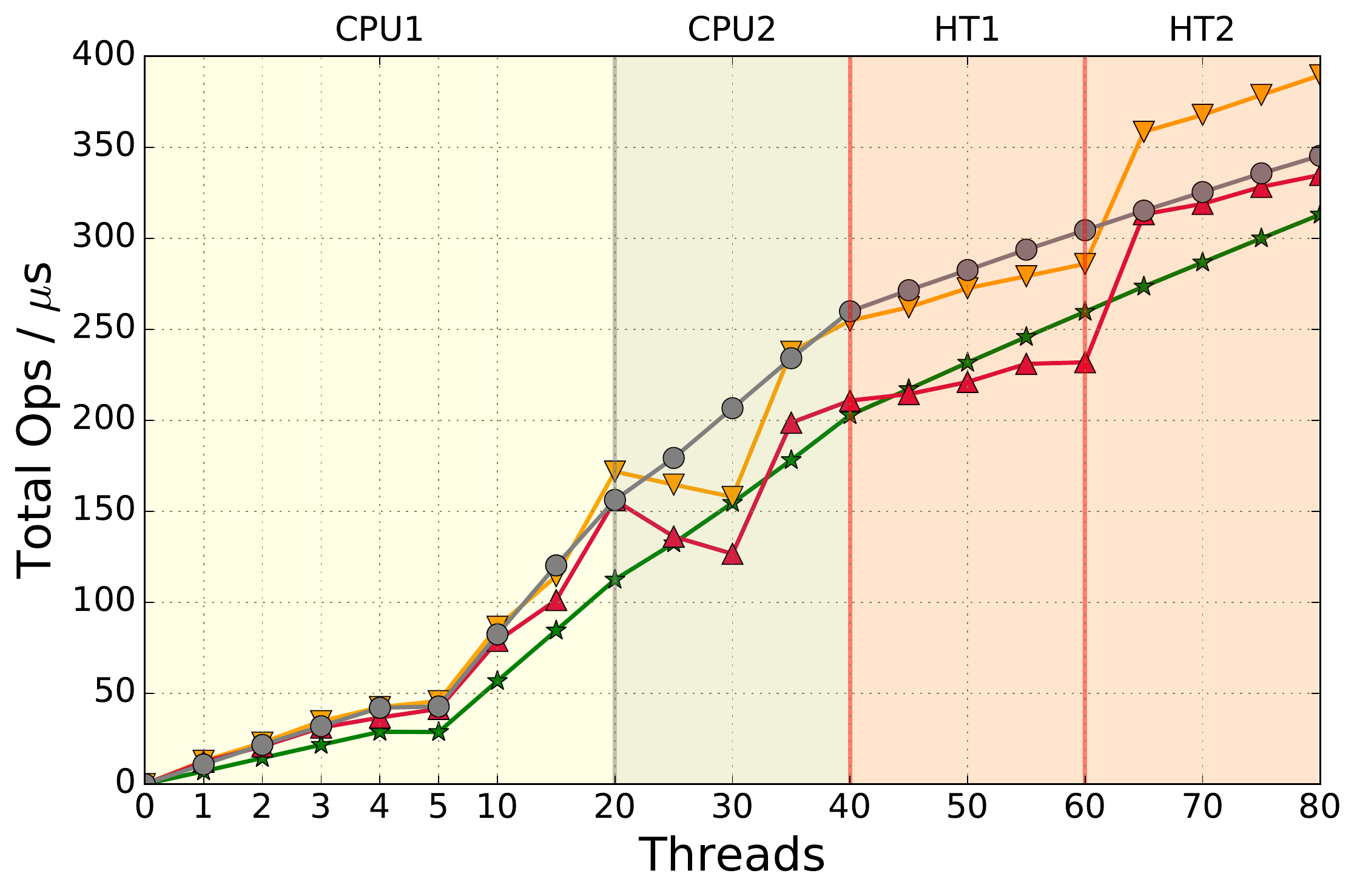}
        \caption{60\% load factor @ 10\%}
        \label{fig:PerfTen60Ichec}
    \end{subfigure}
    \qquad
    \begin{subfigure}[b]{0.38\textwidth}
        \includegraphics[width=\textwidth]{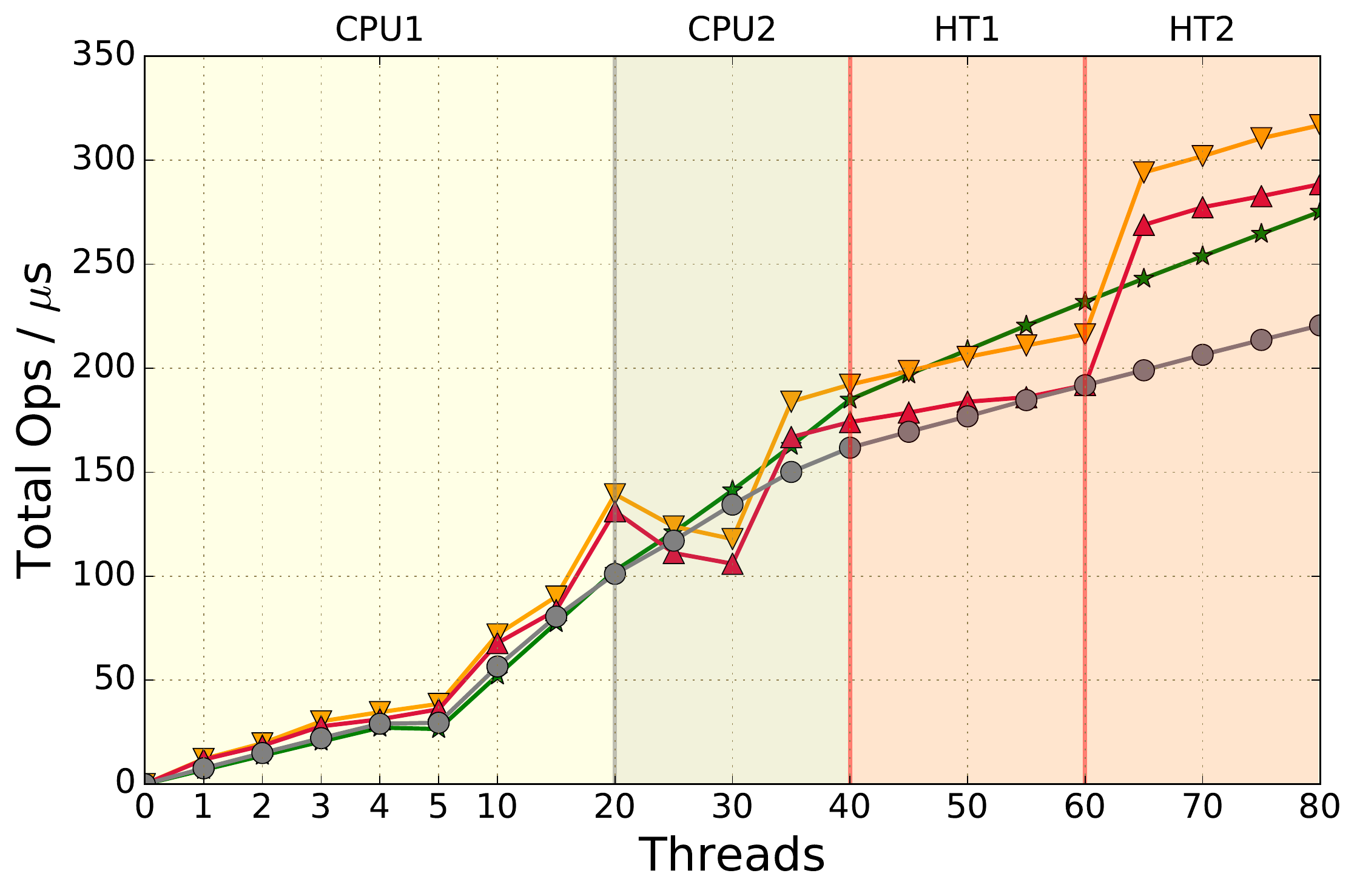}
        \caption{80\% load factor @ 10\%}
        \label{fig:PerfTen80Ichec}
    \end{subfigure}
  
    \begin{subfigure}[b]{0.38\textwidth}
        \includegraphics[width=\textwidth]{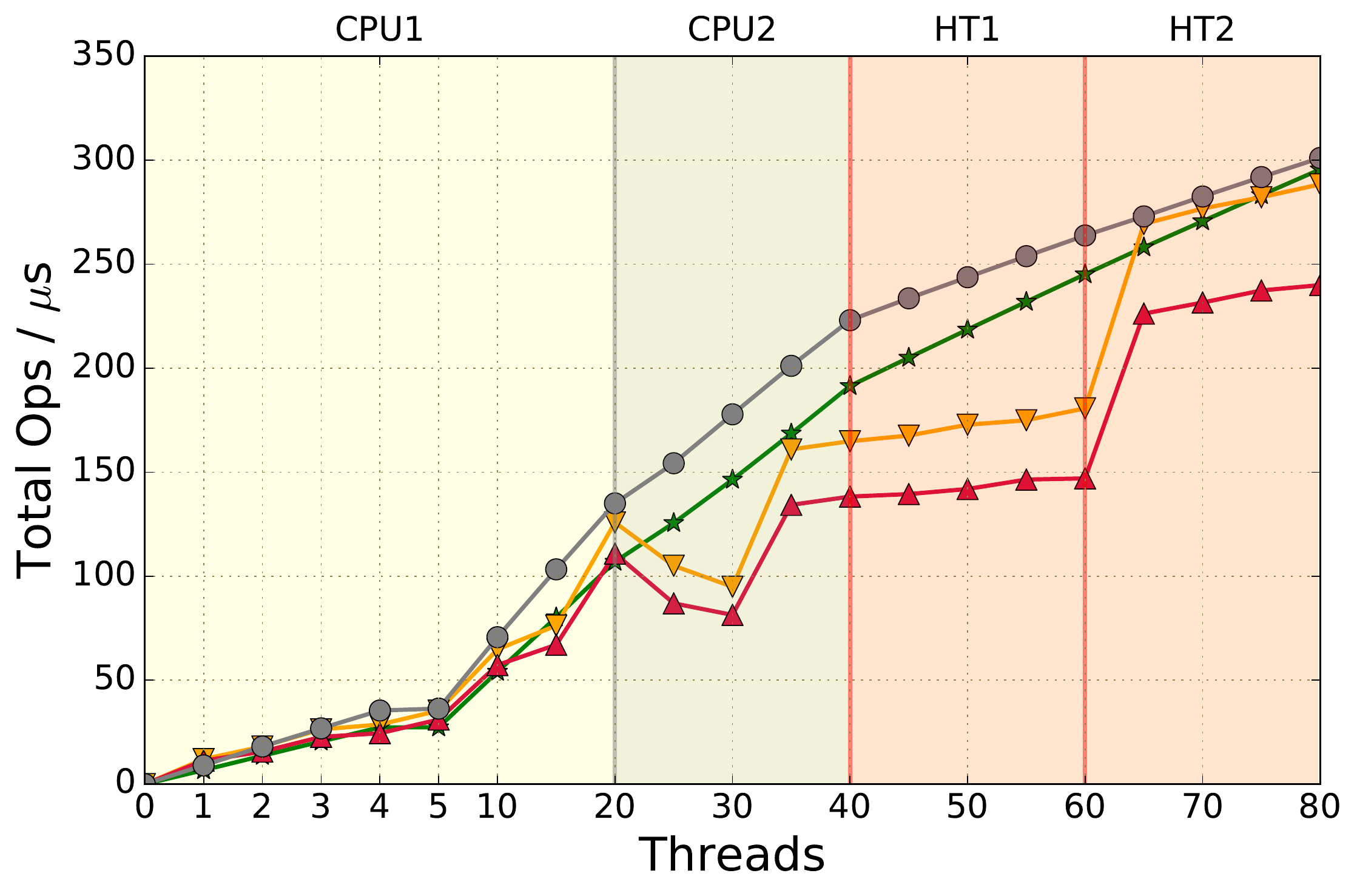}
        \caption{60\% load factor @ 20\%}
        \label{fig:PerfTwenty60Ichec}
    \end{subfigure}
    \qquad
    \begin{subfigure}[b]{0.38\textwidth}
        \includegraphics[width=\textwidth]{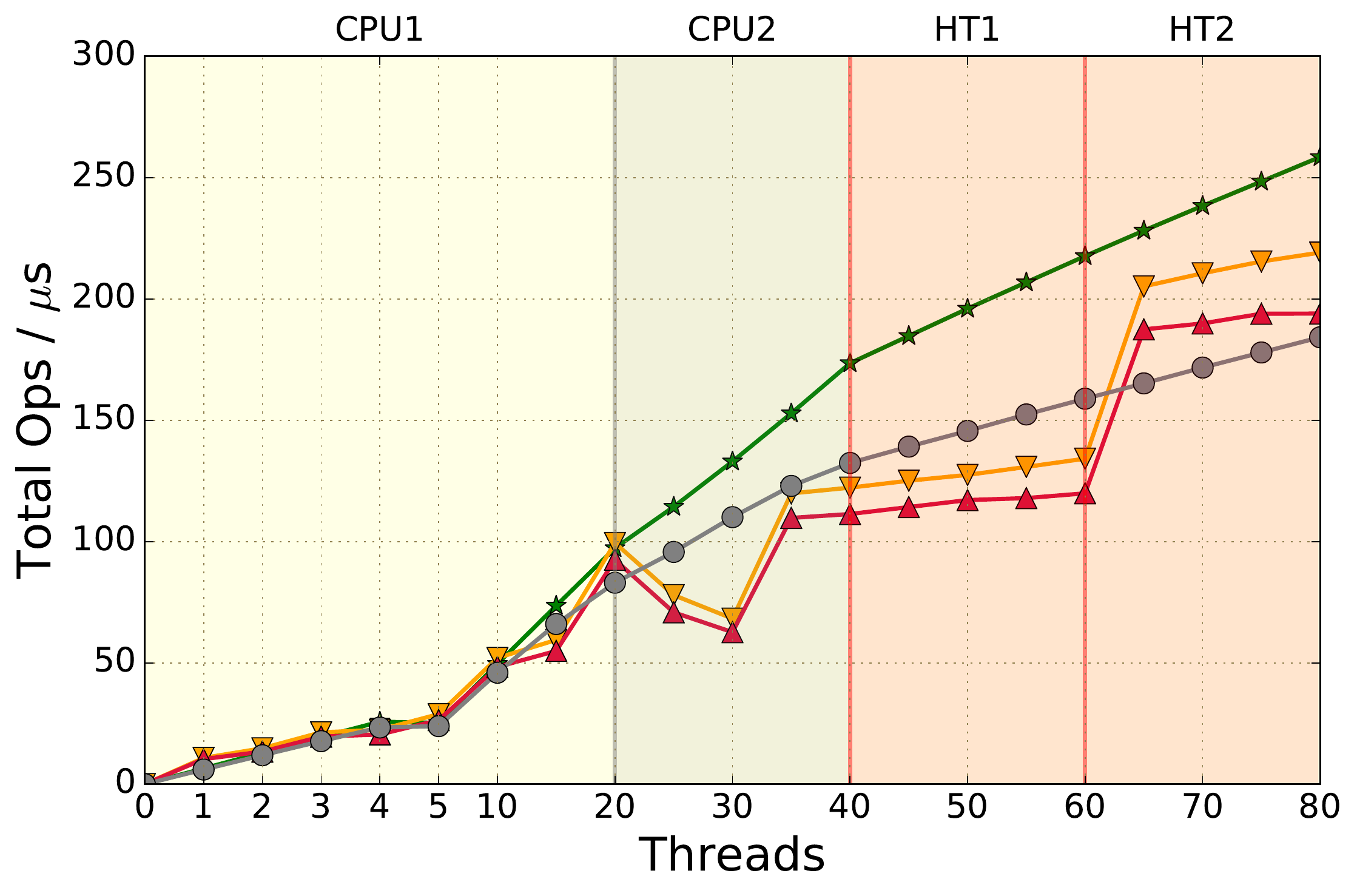}
        \caption{80\% load factor @ 20\%}
        \label{fig:PerfTwenty80Ichec}
    \end{subfigure}
    
    \begin{subfigure}[b]{0.38\textwidth}
        \includegraphics[width=\textwidth]{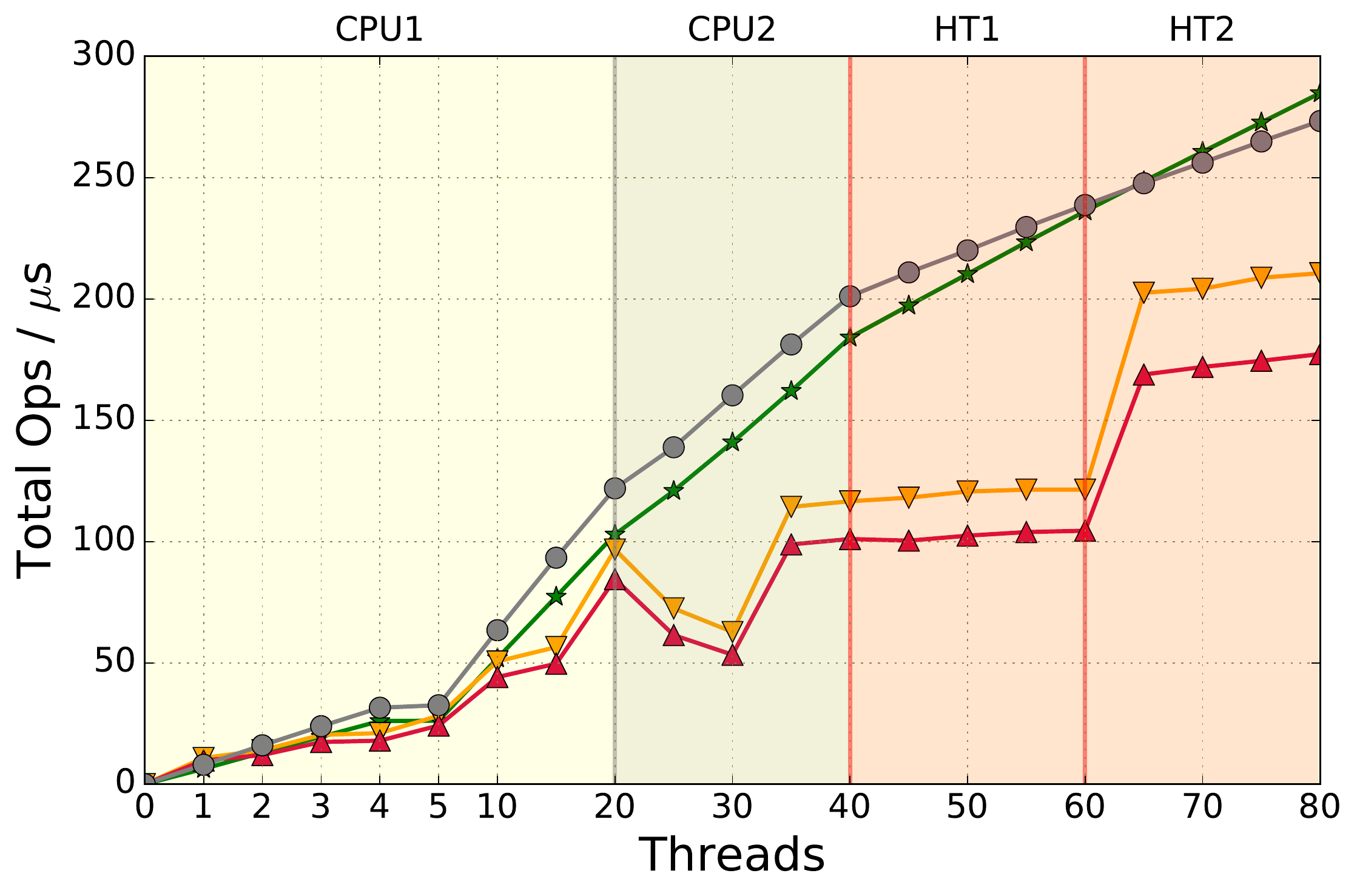}
        \caption{60\% load factor @ 30\%}
        \label{fig:PerfThirty60Ichec}
    \end{subfigure}
    \qquad
    \begin{subfigure}[b]{0.38\textwidth}
        \includegraphics[width=\textwidth]{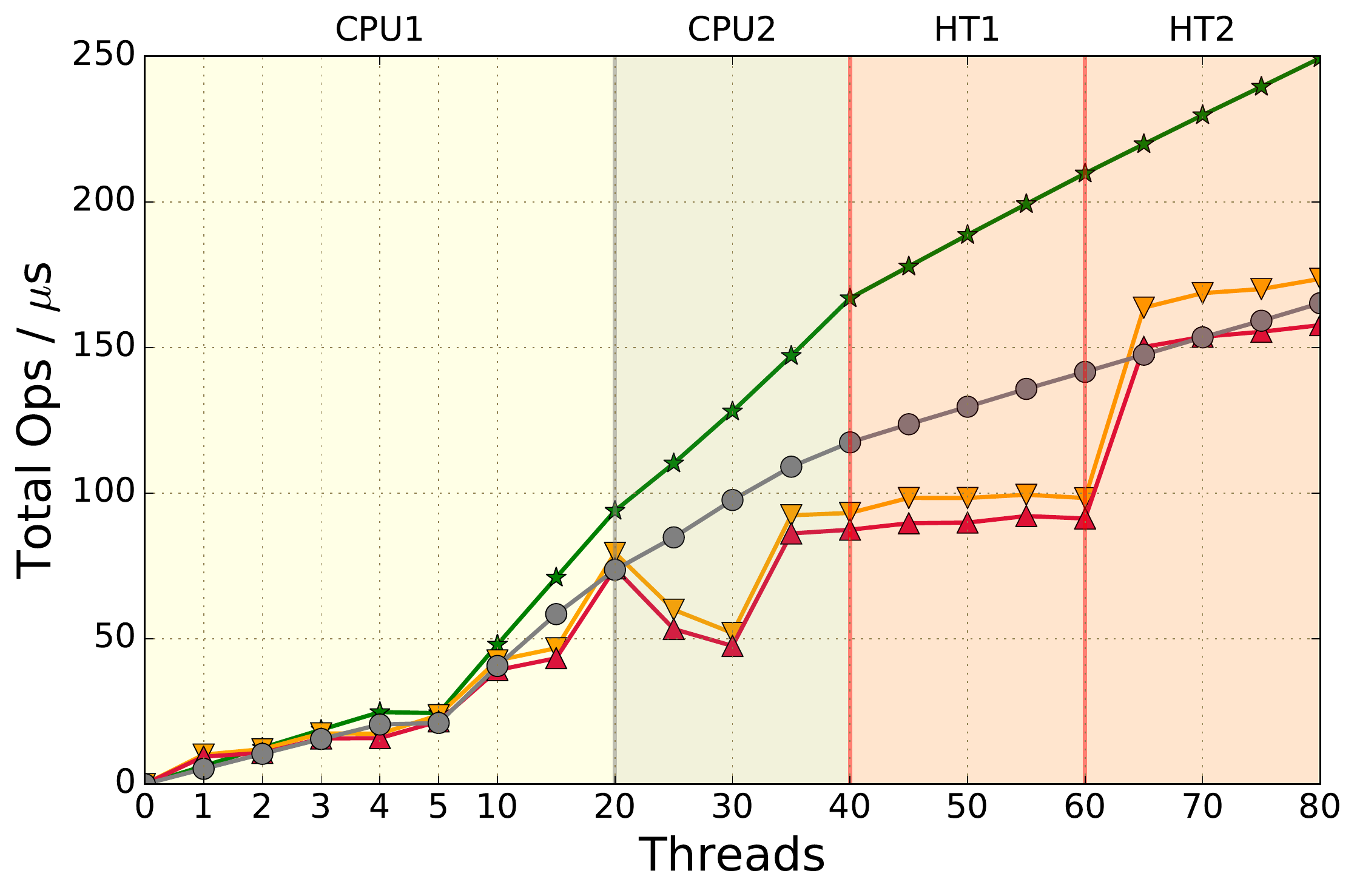}
        \caption{80\% load factor @ 30\%}
        \label{fig:PerfThirty80Ichec}
    \end{subfigure}
    
    \begin{subfigure}[b]{0.38\textwidth}
        \includegraphics[width=\textwidth]{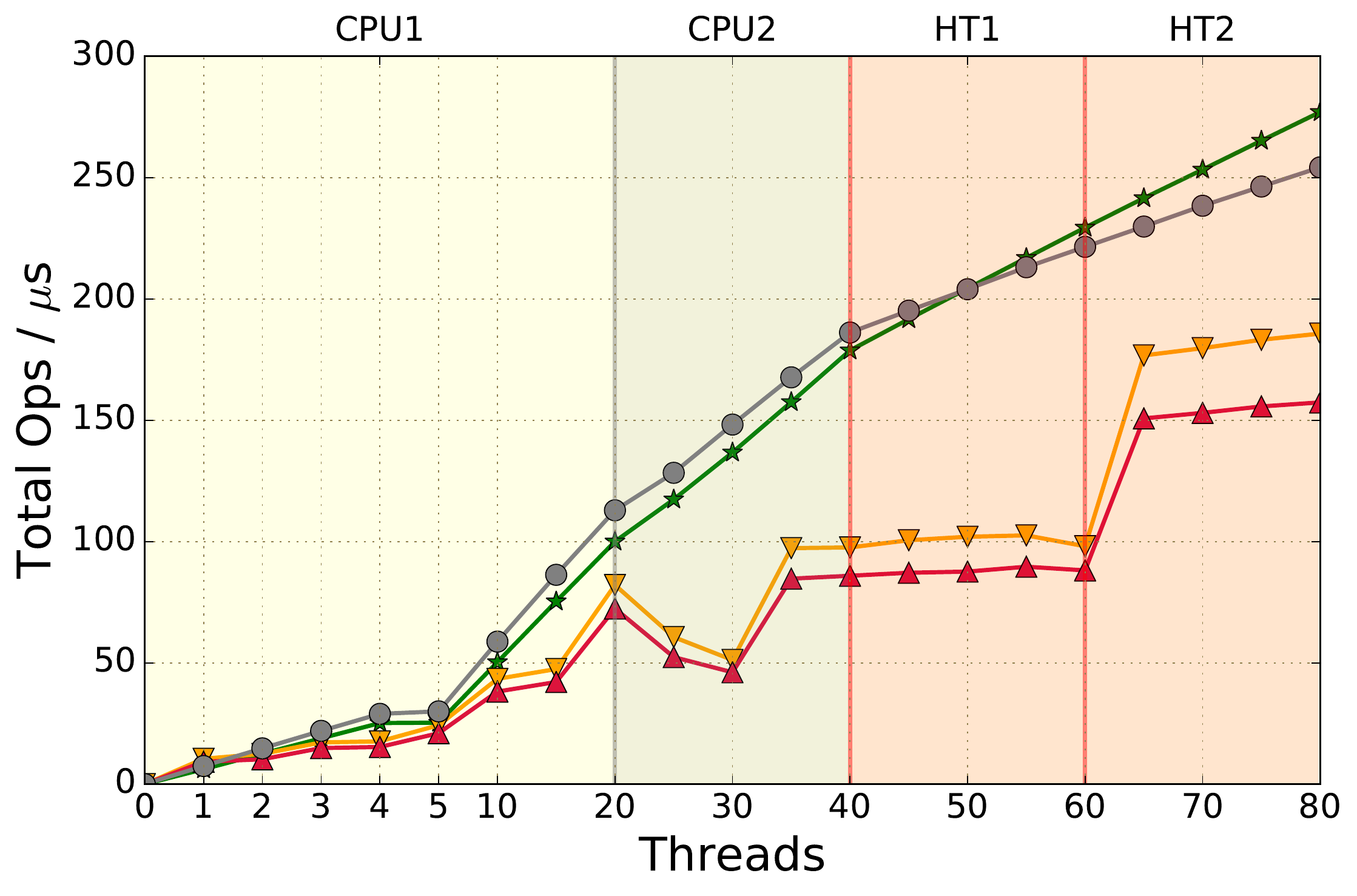}
        \caption{60\% load factor @ 40\%}
        \label{fig:PerfFourty60Ichec}
    \end{subfigure}
    \qquad
    \begin{subfigure}[b]{0.38\textwidth}
        \includegraphics[width=\textwidth]{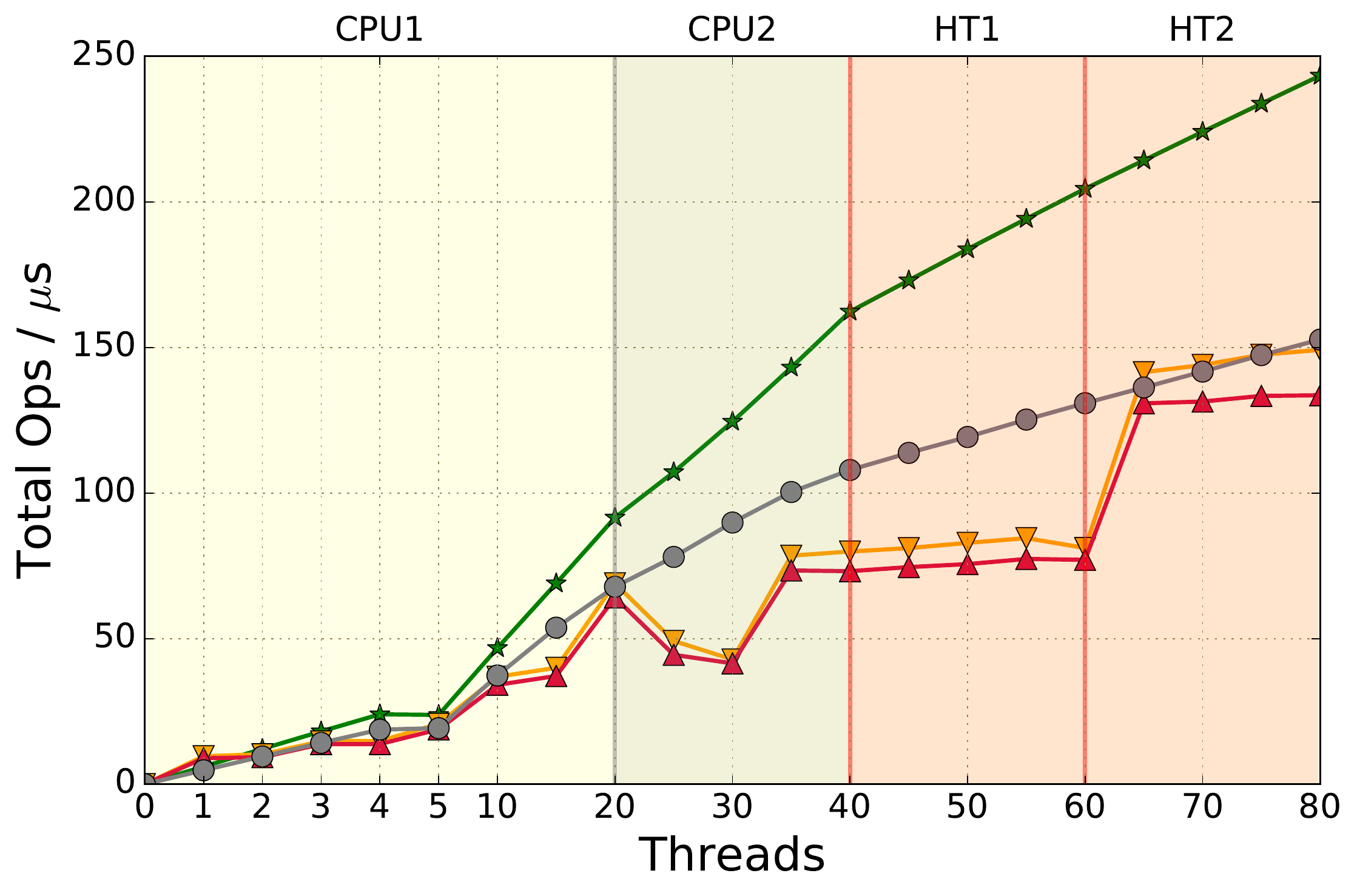}
        \caption{80\% load factor @ 40\%}
        \label{fig:PerfFourty80Ichec}
    \end{subfigure}
    
    \begin{subfigure}[b]{0.8\textwidth}
        \includegraphics[width=\textwidth]{graphs/Hopscotchlegend.pdf}
    \end{subfigure}
    
    \caption{Performance graphs for the second machine.}
    \label{fig:PerfIchecAll}
\end{figure*}

Our testing was performed on another machine to ensure we weren't fitting to a particular hardware architecture during our benchmarking. The machine had 2 CPUs (Intel\textregistered\ Xeon\textregistered\ Gold 6148) with 20 cores each, and 27.5MB of L3 Cache. Each core had two hardware threads, meaning the total number of threads was 80. The machine had 192 GiB of RAM installed. Threads were pinned exactly like our main experiments, except in increments of 5, from 5 to 80. All of our algorithms were compiled with \textit{clang++} 7.0 at \texttt{O3} level of optimisation, and also targeted the specific processor architecture they were being run on. Everything else about the testing process remained the same. The figures for single threaded performance can be seen in Figure \textbf{\ref{fig:SingleThreadIchec}} while the results for throughput as a function of concurrency can be seen in Figure \textbf{\ref{fig:PerfIchecAll}}. The results in Figure \textbf{\ref{fig:PerfIchecAll}} broadly match the same trends as seen in our results above in Figures \textbf{\ref{fig:PerfMitLower}} and  \textbf{\ref{fig:PerfMitHigher}}.

\end{document}